\documentclass[aps,prresearch,reprint,superscriptaddress,showkeys,longbibliography]{revtex4-2}
\usepackage{epsfig}
\usepackage{graphicx}
\usepackage{amssymb}
\usepackage{algorithm}
\usepackage{algorithmic}
\usepackage{upgreek}
\usepackage{mathtools}
\usepackage{bm}
\usepackage{bibentry}

\usepackage{braket}
\usepackage{comment}
\usepackage{tikz}

\usepackage{array}
\usepackage{booktabs}
\usepackage[normalem]{ulem} 
\usepackage{diagbox}
\usepackage{multirow}
\usepackage{physics}
\usepackage{tcolorbox}
\usepackage[dvipsnames]{xcolor}

\definecolor{DarkGreen}{RGB}{0,80,40}
\definecolor{ChangeColor}{RGB}{150,0,0}

\newcommand{\vardbtilde}[1]{\tilde{\raisebox{0pt}[0.85\height]{$\tilde{#1}$}}}

\begin{document}

\title{Exploring fixed points and eigenstates of quantum systems with reinforcement learning}

\author{Mar\'ia Laura Olivera-Atencio} \email{molivera1@us.es}
\affiliation{F\'{\i}sica Te\'orica, Universidad de Sevilla, Apartado 1065, E-41080 Sevilla, Spain}

\author{Jes\'us Casado-Pascual}\email{jcasado@us.es}
\affiliation{F\'{\i}sica Te\'orica, Universidad de Sevilla, Apartado 1065, E-41080 Sevilla, Spain}
\affiliation{Multidisciplinary Unit for Energy Science, Universidad de Sevilla, E-41080 Sevilla, Spain}

\author{Denis Lacroix } \email{lacroix@ijclab.in2p3.fr}
\affiliation{Universit\'e Paris-Saclay, CNRS/IN2P3, IJCLab, 91405 Orsay, France}

\date{\today}
\begin{abstract}
We introduce a reinforcement learning algorithm designed to identify the fixed points of a given quantum operation. The method iteratively constructs the unitary transformation that maps the computational basis onto the basis of fixed points through a reward–penalty scheme based on quantum measurements. In cases where the operation corresponds to a Hamiltonian evolution, this task reduces to determining the Hamiltonian eigenstates. The algorithm is first benchmarked on random Hamiltonians acting on two and three qubits and then applied to many-body systems of up to six qubits, including the transverse-field Ising model and the all-to-all pairing Hamiltonian. In both cases, the algorithm is demonstrated to perform successfully; in the pairing model, it can also reveal hidden symmetries, which can be exploited to restrict learning to specific symmetry sectors. Finally, we discuss the possibility of post-selecting high-fidelity states even when full convergence has not been reached.
\end{abstract}


\maketitle

\section{Introduction}

Quantum machine learning (QML) has emerged as a fertile ground for research in recent years, exploring the interplay between quantum mechanics and machine learning~\cite{NielsenChuang2010, Schuld2015,Torlai_2018, BiamonteNature,Liu_2020,Heese2025}. This field aims to leverage quantum technologies to enhance various machine learning tasks, including classification problems~\cite{Haug2023}, quantum control~\cite{Perrier2020}, or the enhancement of quantum synchronization processes~\cite{Cardenas2019}. Diverse quantum algorithms have been proposed and, in some cases, experimentally implemented, demonstrating potential speedups over classical approaches~\cite{Saini2020,Innan2024}.

Reinforcement learning (RL) represents a fundamental approach within the field of machine learning~\cite{Sutton2018}. In contrast to supervised and unsupervised learning, which mainly focus on identifying patterns in data~\cite{Shrapnel_2018, Agresti_2019, Youssry_2020,Luchnikov_2020, Melnikov_2020}, RL addresses complex problems by framing them as the search for an effective sequence of decisions that maximizes long-term rewards. This is accomplished through the dynamic interaction between an agent and its environment. The learning process is driven by repeated interactions, where actions are either rewarded or penalized based on a predefined policy, aiming to maximize cumulative rewards~\cite{DMDP,Agostinelli2018,Sutton2018}. This iterative trial-and-error methodology, when combined with quantum mechanics, has given rise to a diverse body of work ranging from fundamental studies~\cite{Dong2008}, studies exploring the use of quantum properties to enhance learning tasks~\cite{Paparo2014,Dunjko2016}, to specialized tasks such as preparation of quantum states or the search for quantum error correction strategies~\cite{Bukov_Day2018, Bukov2018, Fosel2018}.

In recent years, this paradigm has continued to inspire further developments in quantum RL~\cite{kaldari2025}. For instance, high fidelity state control has been achieved using quantum RL for optimization when full state observability is infeasible~\cite{Jiang2022}. Policy iteration methods have been explored to improve the performance of quantum reinforcement algorithms in the context of infinite-horizon discounted problems~\cite{Cherrat2023}. Furthermore, challenges such as managing continuous action spaces~\cite{Liu2023}, as well as the implementation of a quantum decision problem in a quantum computer, have been successfully addressed~\cite{Franceschetto2023}, including tackling eigenvalue problems in the presence of dissipation~\cite{OliveraAtencio2023, OliveraAtencio2023_2,OliveraAtencio2025}.

The present work introduces an RL algorithm designed to approximately learn the fixed points of a given quantum operation. More precisely, the proposed algorithm identifies the unitary transformation that maps the computational basis onto the basis composed of the pure states that remain invariant under the action of a specified quantum operation. When the operation corresponds to unitary dynamics generated by a Hamiltonian, the fixed points coincide with its eigenstates, and the algorithm provides the associated eigenbasis. This approach is inspired by previous work~\cite{Albarran_Arriagada_2020}, which focused on identifying the eigenvectors of unknown Hamiltonians. A key difference is that our algorithm learns all fixed points simultaneously, whereas in Ref.~\cite{Albarran_Arriagada_2020} the eigenstates were determined sequentially, one at a time. Instead of identifying individual eigenstates, our method learns the global unitary transformation that connects the computational basis to the fixed-point basis, providing a more compact and parallelizable framework. Moreover, while the previous approach was restricted to unitary processes, the present formulation does not assume unitarity of the quantum operation itself, but requires that its fixed-point set admit an orthonormal basis of pure states spanning the Hilbert space. Unitary dynamics generated by a Hamiltonian constitute the main example of this situation and will be the focus of the present work. However, this structural condition is not exclusive to unitary evolutions: certain non-unitary channels, such as phase-damping noise~\cite{NielsenChuang2010}, also admit orthonormal bases of pure invariant states. Quantum maps whose fixed points are mixed states or do not form an orthonormal basis fall outside the scope of the present unitary-parametrization strategy, and extending the method to such cases would require a different representation of the target manifold. Related RL approaches addressing non-unitary dynamics can be found in Refs.~\cite{OliveraAtencio2023, OliveraAtencio2023_2,OliveraAtencio2025}.

It is worth noting that the problem of determining Hamiltonian eigenstates has been addressed through a wide variety of approaches. Specifically, in recent years, significant efforts have been devoted to the development of hybrid quantum-classical variational techniques aimed at constructing approximate ground states of complex many-body quantum systems~\cite{Mcclean2016,Mcardle2020,Motta2022}. Variational methods based on parametrized trial wavefunctions have been extensively explored in different areas of physics and chemistry~\cite{Cerezo2021,Tilly2022}, offering the advantage, particularly for near-term applications, of being easier to implement on imperfect quantum devices. They have been successfully applied to many-body systems of moderate size; however, as the number of qubits increases, the optimization task typically becomes more challenging and may fail due to the complexity of the energy landscape, with the onset of the so-called ``barren plateau'' phenomenon being a paradigmatic example~\cite{Larocca2025}. Extensive efforts have been made to extend variational approaches and, in some cases, complement them with quantum subspace diagonalization techniques to also provide accurate descriptions of excited states~\cite{Mcclean2017,Grimsley2019,Nakanishi2019,Parrish2019,Higgott2019,Jones2019,Motta2020,Huggins2020,Stair2020,Seki2021,Yordanov2021,Tang2021,Zhang2021b,Yordanov2022,Cortes2022,Epperly2022,Ruizguzman2022,Hlatshwayo22,Wang2023a,Shen2023,Zheng2023,Smart2024,Motta2024,Hirsbrunner2024,Zheng2024,Nakagawa2024,Ding2024,Beaujeault-Taudiere2024,Hlatshwayo2024,Barison2025,Marti-dafcik2025,Grimsley2025,Zhang2025}.  Our aim with the method presented here is not to compete with existing variational algorithms, but rather to explore a non-variational alternative for addressing spectral problems in many-body systems. In particular, the protocol is designed to reconstruct a global invariant basis through adaptive measurement-driven updates, instead of targeting individual eigenstates via energy minimization.
	To illustrate this approach, we apply the RL algorithm to two types of interacting-particle models that are relevant, respectively, for the condensed-matter and nuclear-physics communities.

	The ability to reconstruct the complete invariant basis is not merely a formal aspect of the protocol, but can be relevant in many-body contexts where global spectral information plays a central role. Examples include the analysis of excited-state quantum phase transitions~\cite{Caprio2008}, spectral statistics and level-density estimation in quantum chaotic systems~\cite{Bohigas1984,Haake2010}, as well as studies of thermalization and ergodicity where the structure of the full eigenbasis is relevant~\cite{Srednicki1994,DAlesio2016}.
	In addition, reconstructing the invariant basis provides direct access to symmetry sectors and degeneracy patterns that may not be known a priori, as illustrated in the pairing model discussed later in the manuscript. The present approach should therefore be understood as addressing a problem of global spectral reconstruction rather than that of isolated eigenstate targeting.

 Beyond this difference in learning objective, the RL protocol introduced here is also formulated in a distinct operational setting. Specifically, the protocol assumes the ability to implement the unitary time evolution operator $U(\tau)$ and perform projective measurements, without requiring explicit knowledge of the Hamiltonian operator generating that evolution. This distinguishes the present framework from classical diagonalization and standard variational approaches, which typically rely on explicit operator representations. Such a setting can arise in analog quantum simulators or experimentally implemented many-body systems, where coherent time evolution can be realized and measured even if the microscopic Hamiltonian is not known in closed form~\cite{Georgescu2014}. A similar situation occurs in periodically driven (Floquet) systems, where one may have experimental access to the stroboscopic evolution operator over one driving period, while the corresponding effective Floquet Hamiltonian is not available as an explicit matrix representation~\cite{Bukov2015,Eckardt2017}. In these contexts, learning spectral properties directly from dynamical evolution can be operationally meaningful.
 
 It is worth emphasizing that the present implementation should be regarded as a proof-of-principle approach, as the computational cost grows exponentially with the Hilbert-space dimension (see Sec.~\ref{subsec:scaling}), and scalability to large many-body systems is therefore not established. Despite this limitation, the interest of the method lies in the type of learning task it addresses and the operational setting in which it is formulated, as discussed above. In particular, the approach enables the identification of invariant structures of quantum dynamics in a global manner, does not require explicit knowledge of the underlying Hamiltonian but only access to the corresponding evolution, and may be applied to classes of quantum operations beyond strictly unitary dynamics when an orthonormal invariant basis exists. It also provides direct access to symmetry sectors and degeneracy structures through the learned invariant basis (see Secs.~\ref{subsec:pairing} and \ref{subsec:sym_rest_RL}). From a broader perspective, this measurement-driven and reinforcement-learning-based formulation offers an alternative viewpoint for analyzing spectral properties of quantum systems, and may provide a basis for further methodological developments.
 		
The remainder of the paper is organized as follows. In Sec.~\ref{sec:algo}, we provide a detailed description of the algorithm. We first define the problem statement and goal, specifying the task of learning the fixed points of a quantum operation. We then present the RL protocol step by step, detailing the iterative procedure and the reward–punishment strategy.  This is followed by a formulation of the algorithm within the RL framework, together with a pseudocode description provided in Appendix~\ref{appendix}, and a discussion of resource considerations and scaling. Next, we discuss strategies to prevent convergence to spurious fixed points and describe a fine-tuning procedure after initial convergence to further improve fidelities. Metrics for assessing convergence and accuracy are also introduced, followed by initial benchmarks on random Hamiltonians that illustrate the performance of the method, and an analysis of hyperparameter selection and robustness. Section~\ref{sec:applications} focuses on applications to physical systems, namely the transverse-field Ising model (TFIM) and the pairing model, and discusses extensions such as symmetry-restricted RL and post-selection of high-fidelity states. Finally, Sec.~\ref{sec:conclusions} summarizes our main findings and outlines possible directions for future work.

\section{Algorithm description}
\label{sec:algo}
\subsection{Problem statement and goal}
\label{subsec:algosta}
Consider a quantum system with a state space of dimension $d$ and computational basis $\mathsf{C} = \{\ket{j}\}_{j \in \{0,\dots,d-1\}}$. Let  $\mathsf{B} = \{\ket{\Phi_{\alpha}}\}_{\alpha \in \{0,\dots,d-1\}}$ be another, unknown orthonormal   target  basis that also spans the state space of the system. We assume that the only information available about the basis $\mathsf{B}$ is that it is left invariant by a quantum operation $\mathcal{E}$, meaning that
$\mathcal{E}(\ketbra{\Phi_{\alpha}}{\Phi_{\alpha}}) = \ketbra{\Phi_{\alpha}}{\Phi_{\alpha}}$ for all $\alpha \in \{0,\dots,d-1\}$. With this hypothesis our objective is to construct, at least approximately, a unitary transformation that maps the computational basis $\mathsf{C}$ onto the target basis $\mathsf{B}$ (up to a permutation of their elements). 

A particular case---on which we will focus later in this work---is when the target basis $\mathsf{B}$ consists of the (unknown) stationary states of a certain Hamiltonian $H$, that is, $H\ket{\Phi_{\alpha}} = E_{\alpha} \ket{\Phi_{\alpha}}$ for all $\alpha \in \{0, \dots, d-1\}$, where $E_{\alpha}$ is the eigenenergy associated with the stationary state $\ket{\Phi_{\alpha}}$. In this setting, the target basis $\mathsf{B}$ remains invariant under the action of the quantum operation  
\begin{equation}
	\label{evolution}
	\mathcal{E}(\bullet) = U(\tau) \bullet U^\dagger(\tau),
\end{equation}
where $U(\tau) = e^{-i \tau H / \hbar}$ is the unitary time-evolution operator from time $0$ to $\tau$, with $\tau$ being a time interval whose value is, for now, arbitrary.
As we will see later, the particular choice of the parameter $\tau$ plays a crucial role in the correct functioning of the algorithm.

The goal of our quantum RL method is to iteratively construct a sequence of unitary matrices $\{D_k\}$, where $k$ labels the iteration number, such that 
\begin{eqnarray}
	\lim_{k \rightarrow +\infty} D_k \ket{j}  \approx \ket{ \Phi_\alpha} \label{eq:limitDk}
\end{eqnarray}
for all $\ket{j}$, with $\ket{\Phi_\alpha}$ representing some state of the basis $\mathsf{B}$. In other words, in the limit $k \to +\infty$, $D_k$ approximately maps the basis $\mathsf{C}$ onto the basis $\mathsf{B}$, up to a certain permutation of its vectors.

\subsection{Technical details on the iterative process}
\label{subsec:algotech}
The unitary transformation generated in the $(k+1)$th iteration, $D_{k+1}$, is obtained from the unitary operator $D_k$, corresponding to iteration $k$, according to the update rule
\begin{equation}
	\label{ITD}	
	D_{k+1}=D_k \prod_{j=0}^{d-2}\prod_{l=j+1}^{d-1}D_k^{(j,l)},
\end{equation}
where the $d(d-1)/2$ operators $D_k^{(j,l)}$ are unitary transformations.  
The index $(j,l)$ refers to the $\mathsf{C}$ basis states. In this product, the operators are arranged from left to right according to the increasing sequence of index pairs $(j, l)$, ordered lexicographically. Specifically, $(j_1, l_1)$ precedes $(j_2, l_2)$ if either $j_1 < j_2$, or $j_1 = j_2$ and $l_1 < l_2$. 

We assume that the unitary operators $D_k^{(j,l)}$ are taken as rotations acting on the two-dimensional subspace spanned by $\ket{j}$ and $\ket{l}$. These rotations can be expressed in terms of the Pauli-type operators
\begin{eqnarray}
	\left\{ 
	\begin{array}{l}
		X^{(j,l)} = \ketbra{j}{l} + \ketbra{l}{j} , \\
		\\
		Y^{(j,l)} = -i (\ketbra{j}{l} - \ketbra{l}{j}) ,\\
		\\
		Z^{(j,l)} = \ketbra{j}{j} - \ketbra{l}{l},
	\end{array}
	\right.
\end{eqnarray}
which serve as their generators.  Specifically, the operators $D_k^{(j,l)}$ are expressed as
\begin{equation} 
\label{eq:random-rot}
D_k^{(j,l)}= e^{-i \beta_k^{(j,l)} Y^{(j,l)}/2} e^{-i \gamma_k^{(j,l)} Z^{(j,l)}/2} e^{-i \alpha_k^{(j,l)} X^{(j,l)}/2}, 
\end{equation}
where the angles $\alpha_k^{(j,l)}$, $\beta_k^{(j,l)}$, and $\gamma_k^{(j,l)}$ are chosen depending on whether the corresponding action is rewarded or penalized. In the case of a reward, all angles are set to zero, so that $D_k^{(j,l)}$ reduces to the identity operator $I$. This corresponds to no exploration, i.e., pure exploitation in RL terms. Conversely, in the case of a penalty, the angles are drawn pseudo-randomly and uniformly within intervals that are dynamically adjusted during the learning process, allowing exploration of the space of unitary operators.

Schematically, these intervals are expanded in response to penalties and contracted in response to rewards, thereby regulating  the extent of exploration in the space of unitary operators. The degree of exploration at iteration $k$ is controlled via a set of $d(d-1)/2$ exploration parameters $w_k^{(j,l)} \leq 1$, whose evolution with the number of iterations depends on the cumulative number of rewards and punishments received up to that point (see the discussion below for further details).

\subsection{RL algorithm step-by-step}
\label{subsec:algostep}
The algorithm is implemented on a set of $d$ $d$-dimensional systems (qudits), each initialized in a distinct state of the computational basis. Specifically, the initial density operator of the full system is
\begin{equation}
	\rho_1=\ketbra{0}{0} \otimes \ketbra{1}{1} \otimes \dots \otimes  \ketbra{d-1}{d-1}.
\end{equation}
We note that the use of $d$ parallel qudit registers is not essential. The same procedure can be implemented sequentially on a single register by preparing each computational basis state $\ketbra{j}{j}$ in turn and applying the algorithmic steps independently. No inter-qudit entanglement or collective measurement is required. In practice, such a sequential implementation requires only a single register together with repeated state preparation and measurement. The parallel formulation adopted here is therefore primarily conceptual, as it emphasizes that the objective of the algorithm is to learn a single global unitary transformation that consistently maps the entire computational basis onto the invariant basis.
	Although the experimental executions associated with different basis states may be time-multiplexed, the learning dynamics remain intrinsically global. In particular, all updates contribute to the construction of the same unitary transformation $D_k$, whose structure is constrained both by unitarity and by consistency across different basis states.

The iterative process begins with the initial assignments $D_1 = I$ and $w_1^{(j,l)} = 1$ for all index pairs $(j,l)$ such that $0 \leq j < l \leq d-1$. At each iteration, the values of $D_{k+1}$ and the parameters $w_{k+1}^{(j,l)}$ are updated based on those from the previous iteration, $D_k$ and $w_{k}^{(j,l)}$, following  the procedure described below:

1.  \textbf{Apply $\boldsymbol{D_k}$:}  The unitary transformation $D_k$ is applied to each qudit individually, yielding the density operator
\begin{equation}
	\rho_k = \rho_{k}^{(0)} \otimes  \rho_{k}^{(1)} \otimes \dots \otimes  \rho_{k}^{(d-1)},
\end{equation}
with $\rho_{k}^{(j)}=D_k \ketbra{j}{j} D_k^{\dagger}$.

2. \textbf{Apply $\boldsymbol{\mathcal{E}}$:}  Next, the quantum operation $\mathcal{E}$ is applied to each qudit individually, resulting in the updated density operator
\begin{equation}
	\tilde{{\rho}}_k = \tilde{\rho}_k^{(0)} \otimes \tilde{\rho}_k^{(1)} \otimes \dots \otimes \tilde{\rho}_k^{(d-1)},
\end{equation}
with $\tilde{\rho}_{k}^{(j)}=\mathcal{E}(\rho_{k}^{(j)})$.

3. \textbf{Undo $\boldsymbol{D_k}$:} The next step consists in reversing the unitary transformation applied in step 2 for each qudit. This yields the density operator
\begin{equation}
	\vardbtilde{\rho}_k=\vardbtilde{\rho}_k^{(0)}\otimes \vardbtilde{\rho}_k^{(1)}\otimes \dots \otimes \vardbtilde{\rho}_k^{(d-1)},
\end{equation}
with $\vardbtilde{\rho}_k^{(j)}=D_k^{\dagger} \tilde{\rho}_{k}^{(j)} D_k$. 

4. \textbf{Measure:} Then, a projective measurement is performed on each qudit in the computational basis, with the result for the $j$th qudit denoted by $m_k^{(j)} \in \{0, 1, \dots, d-1\}$. It is worth noting that if, at a given iteration $k$, the target were exactly reached---namely, if $D_k$ mapped the computational basis $\mathsf{C}$ onto the basis $\mathsf{B}$ (up to a permutation of their elements)---then the density operator $\rho_k$ would remain invariant under the quantum operation $\mathcal{E}$. In that case, we would have $\tilde{\rho}_k = \rho_k$, and upon reversing the transformation $D_k$, the initial density operator would be recovered, i.e., $\tilde{\tilde{\rho}}_k = \rho_1$.  If this condition held, the measurement outcomes would satisfy the equality $m_k^{(j)} = j$ for all $j \in \{0, \dots, d-1\}$ with probability one. This is equivalent to stating that, for each label $j$ and any $l \ne j$, the condition $m_k^{(j)} \ne l$ would hold with probability one.  Any outcome consistent with this condition would indicate that the goal may have been achieved and should therefore be rewarded, whereas any violation would imply that the objective has not yet been reached and should be penalized.

5. \textbf{Determination of  $\boldsymbol{D_k^{(j,l)}}$ and $\boldsymbol{w_{k+1}^{(j,l)}}$ (reward/punishment):} Depending on the outcomes of the previous measurements, the following procedure is carried out for each pair of qudits $(j, l)$ with $0 \leq j < l \leq d-1$:

i) \textbf{Both correct  (double reward):} If $m_k^{(j)} \neq l$ and $m_k^{(l)} \neq j$, the condition is satisfied for both qudits. In this case, a double reward is applied by reducing the corresponding exploration parameter according to the rule 
\begin{equation}
\label{doublereward}
w_{k+1}^{(j,l)} = r^2 w_k^{(j,l)},
\end{equation}
 where $r \in (0,1)$ is a parameter referred to as the reward rate. Furthermore, the three angles in Eq.~(\ref{eq:random-rot}) are set to zero, so that $D_k^{(j,l)}$ reduces to the identity operator, i.e., $D_k^{(j,l)} = I$.

ii) \textbf{One correct and one incorrect (reward $\boldsymbol{\&}$ punishment):} If $m_k^{(j)} \neq l$ and $m_k^{(l)} = j$, or $m_k^{(j)} = l$ and $m_k^{(l)} \neq j$, then one qudit satisfies the condition while the other violates it. In this case, a mixed reward–punishment update is performed. Specifically,  the exploration parameter is updated according to the rule 
\begin{equation}
\label{rewardpunishment}
w_{k+1}^{(j,l)} = \min (r p  w_k^{(j,l)},1),
\end{equation}
 where $p > 1$ is a parameter referred to as the punishment rate. Thus, the exploration parameter decreases if $rp < 1$, remains unchanged if $rp = 1$, and increases if $rp > 1$, with the $\min$ function preventing it from exceeding $1$. The unitary operator $D_k^{(j,l)}$ is then constructed as in Eq.~(\ref{eq:random-rot}), with the angles $\alpha_k^{(j,l)}$, $\beta_k^{(j,l)}$, and $\gamma_k^{(j,l)}$ drawn pseudo-randomly and uniformly within the interval $[-\pi w_k^{(j,l)}, \pi w_k^{(j,l)}]$.

iii) \textbf{Both incorrect  (double punishment):} If $m_k^{(j)} = l$ and $m_k^{(l)} = j$, then both qudits violate the condition. In this case, a double punishment is applied, with the exploration parameter updated according to the rule
\begin{equation}
\label{doublepunishment}
w_{k+1}^{(j,l)} = \min(p^2 w_k^{(j,l)},1).
\end{equation}
The construction of $D_k^{(j,l)}$ follows the same procedure as in the case ii).

6. \textbf{Determination of $\boldsymbol{D_{k+1}}$:} Once the above procedure has been applied to compute the operators $D_k^{(j,l)}$ for all qudit pairs, the operator $D_{k+1}$ is obtained from $D_k$ using Eq.~(\ref{ITD}).

7. \textbf{Reset:} If the measurement outcome $m_k^{(j)}$ differs from $j$, the unitary $\ketbra*{m_k^{(j)}}{j}+\ketbra*{j}{m_k^{(j)}}+\sum_{l\neq j, m_k^{(j)}}\ketbra{l}{l}$ is applied to the corresponding qudit to reset it to its initial computational basis state. Once all necessary qudits have been reinitialized in this way, the algorithm resumes from step 1 with the iteration index updated to $k+1$.

\begin{figure}[htbp]
	\includegraphics[width=0.485\textwidth]{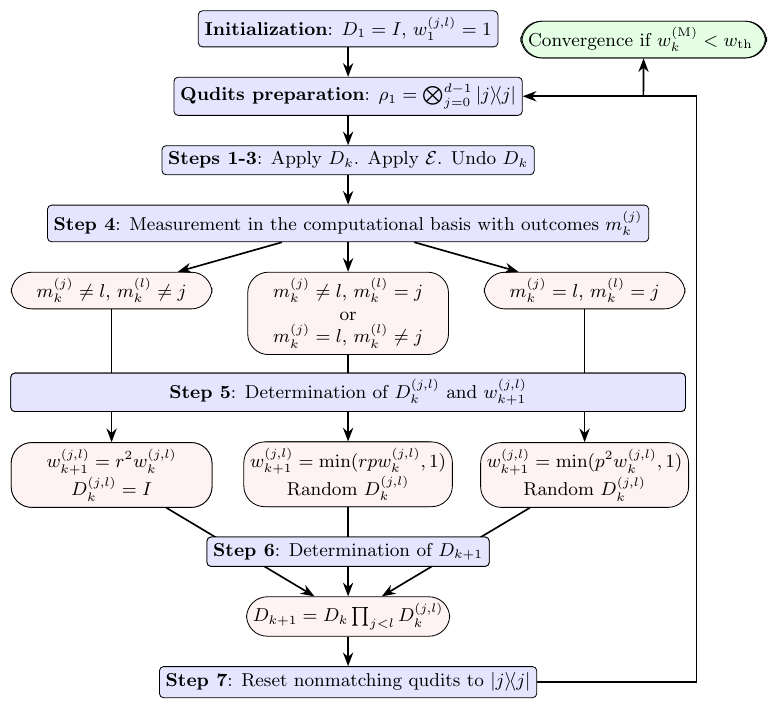}
	\caption{Schematic representation of the proposed RL algorithm.
	}
	\label{fig1}
\end{figure}

The algorithm described above is said to converge if all exploration parameters $w_k^{(j,l)}$ approach zero as the number of iterations $k$ becomes sufficiently large. In this case, all the operators $D_k^{(j,l)}$ tend to the identity, and the unitary transformation $D_k$ approaches a fixed value. To determine in practice whether the algorithm has converged, it is useful to introduce a sufficiently small threshold value for the exploration parameter, denoted by $w_{\mathrm{th}}$. Additionally, we define the maximum exploration parameter at iteration $k$ as
\begin{equation}
	w_k^{(\mathrm{M})} = \max\limits_{0 \leq j < l \leq d-1}  w_k^{(j,l)}.\label{eq:max}
\end{equation}
We say that the algorithm has converged if $w_k^{(\mathrm{M})} < w_{\mathrm{th}}$ after a certain number of iterations $k$. A schematic representation of the algorithm described above is shown in Fig.~\ref{fig1}.

\subsection{RL formulation}

The protocol described in Sec.~\ref{subsec:algostep} can be formulated naturally within the framework of RL~\cite{Sutton2018}. More precisely, it corresponds to a stochastic policy-search scheme of bandit type with adaptive exploration~\cite{Sutton2018,LattimoreSzepesvari}. The algorithm does not rely on a value function, temporal-difference learning, or long-horizon return estimation. Instead, actions are evaluated through immediate reward signals that directly update the exploration parameters.

In this interpretation, the adaptive controller that selects the rotation parameters defining the unitary transformation $D_k$ plays the role of the learning agent. The environment is defined by the fixed quantum operation $\mathcal{E}$ together with projective measurements in the computational basis. The quantum operation itself remains fixed throughout the learning process and therefore the environment does not evolve dynamically in response to the agent's actions.

At iteration $k$, the internal state of the agent is given by the current unitary transformation $D_k$ together with the set of exploration parameters $\{w_k^{(j,l)}: 0 \leq j  < l\leq d-1 \}$. These variables determine the stochastic policy used to select subsequent actions.

For a given pair $(j,l)$, an action consists of selecting a two-dimensional unitary rotation $D_k^{(j,l)}$, parametrized by the angles $\alpha_k^{(j,l)}$, $\beta_k^{(j,l)}$, and $\gamma_k^{(j,l)}$ [see Eq.~\eqref{eq:random-rot}]. The global transformation is updated by composing all pairwise rotations according to Eq.~\eqref{ITD}. The bandit character of the protocol follows from the fact that, for each pair $(j,l)$, an action is selected, a reward is obtained immediately from the measurement outcomes, and the corresponding exploration parameter is updated.

Operationally, after applying $D_k$, the quantum operation $\mathcal{E}$, and undoing $D_k$, projective measurements yield outcomes $m_k^{(j)}$ for all $j$. For each pair $(j,l)$, a binary reward signal is assigned according to whether the condition $m_k^{(j)} \neq l$ and $m_k^{(l)} \neq j$ is satisfied, leading to the multiplicative update rules specified in Eqs.~\eqref{doublereward}--\eqref{doublepunishment}.

The policy is stochastic and explicitly defined as follows. Whenever exploration is required for a given pair $(j,l)$, the rotation angles are sampled independently from uniform distributions centered at zero with half-width $\pi w_k^{(j,l)}$. The exploration parameter $w_k^{(j,l)} \in (0,1]$ therefore directly controls the width of the action distribution: larger values allow broader exploration of the corresponding two-dimensional unitary manifold, whereas successive rewards reduce $w_k^{(j,l)}$ multiplicatively and progressively restrict the search to a neighborhood of the identity.

A complete step-by-step formulation of the protocol, including all update rules in algorithmic form, is provided in Appendix~\ref{appendix}.

\subsection{Resource considerations and scaling}
\label{subsec:scaling}

The resource requirements of the protocol can be estimated as follows. At each iteration, the algorithm updates all pairs of computational-basis indices $(j,l)$ with $0 \le j < l \le d-1$. The number of such pairs is $d(d-1)/2$, implying $O(d^2)$ two-dimensional subspace rotations per iteration.

In a sequential implementation, the protocol requires preparing each computational basis state $\ketbra{j}{j}$ and applying the quantum operation $\mathcal{E}$ once per iteration. This results in $O(d)$ calls to $\mathcal{E}$ and $O(d)$ projective measurements per iteration. As noted above, the $d$-register formulation is purely conceptual and does not require $d$ physical copies of the system; a single register reused sequentially suffices.

For an $N$-qubit system, where $d=2^N$, the number of subspace updates therefore scales as $O(d^2)$ per iteration, while the number of calls to $\mathcal{E}$ and measurements scale as $O(d)$ per iteration. This scaling follows directly from the number of two-dimensional subspaces and from the need to probe each computational basis state once per iteration.

In practice, as convergence is approached, an increasing number of two-dimensional rotations become effectively trivial (i.e., close to the identity), since their associated subspaces already satisfy the invariance condition. Consequently, the number of nontrivial updates typically decreases during later iterations. This behavior does not modify the asymptotic scaling but can reduce the effective computational effort near convergence.

In numerical simulations, the exponential growth of the Hilbert-space dimension makes the classical simulation of the protocol increasingly demanding. In particular, both the quantum evolution and the associated projective measurements must be simulated at each iteration, together with the stochastic updates of the two-dimensional rotations. The present implementation should therefore be regarded as a proof-of-principle approach suitable for moderate Hilbert-space dimensions. When symmetries are present, restricting the learning dynamics to symmetry sectors---as illustrated later in the pairing-model example---can significantly reduce the effective Hilbert-space dimension explored by the algorithm.

\subsection{Fixed-point ambiguities: ensuring convergence to the target basis}
\label{subsec:algofpa}
As noted in Sec.~\ref{subsec:algosta}, the purpose of the algorithm is to construct a unitary transformation that maps the computational basis $\mathsf{C}$ onto the target basis $\mathsf{B}$, under the assumption that $\mathsf{B}$ consists of pure states left invariant by a certain quantum operation $\mathcal{E}$. If, however, $\mathsf{B}$ is not the only invariant set of pure states under such an operation, the algorithm may instead converge to a unitary mapping $\mathsf{C}$ onto a different invariant set, rather than the intended target basis.

To illustrate this point, let us consider the case where the target basis is formed by the stationary states of a Hamiltonian $H$, and the quantum operation $\mathcal{E}$ is given by Eq.~(\ref{evolution}). In this situation, it is straightforward to verify that, for certain values of the evolution time $\tau$, pure states other than the stationary states of $H$ may also remain invariant under $\mathcal{E}$. A necessary and sufficient condition for a pure state $\ketbra{\Psi}{\Psi}$ to be invariant is that the survival probability after evolution for a time $\tau$, defined as
\begin{equation}
\label{survival_probability1}
P=\expval{U(\tau)\ketbra{\Psi}{\Psi}U^{\dagger}(\tau)}{\Psi}=\abs{\expval{U(\tau)}{\Psi}}^2,
\end{equation}
equals unity. Expanding $\ket{\Psi}$ in the eigenbasis of $H$ as $\ket{\Psi} = \sum_{\alpha=0}^{d-1} c_{\alpha} \ket{\Phi_{\alpha}}$, one finds
\begin{equation}
	\label{survival_probability2}
	P=1-4 	\sum_{\alpha=0}^{d-2}\sum_{\beta=\alpha+1}^{d-1}\abs{c_\alpha}^2\abs{c_\beta}^2 \left[\sin\left(\frac{\omega_{\alpha,\beta}\tau}{2}\right)\right]^2,
\end{equation}
with $\omega_{\alpha,\beta} = (E_{\alpha} - E_{\beta})/\hbar$. According to this expression, if $\tau$ is such that $\omega_{\alpha,\beta} \tau$ is an integer multiple of $2\pi$, then $\sin(\omega_{\alpha,\beta}\tau/2)=0$ and any superposition of the form $\ket{\Psi} = c_{\alpha}\ket{\Phi_{\alpha}} + c_{\beta}\ket{\Phi_{\beta}}$ remains invariant under $\mathcal{E}$.
However, unless $\omega_{\alpha,\beta}=0$ (degenerate eigenvalues), such states are not true eigenstates of $H$ when both coefficients $c_{\alpha}$ and $c_{\beta}$  are nonzero.

This situation leads to an ambiguity in the learning dynamics: For certain values of $\tau$, the algorithm cannot reliably distinguish these superpositions from the true eigenstates based solely on their invariance under the quantum operation in Eq.~(\ref{evolution}). Moreover, the issue is not limited to cases where $\omega_{\alpha,\beta} \tau$ is exactly resonant, i.e., an integer multiple of $2\pi$; even when $\omega_{\alpha,\beta} \tau$ takes near-resonant values, the algorithm may struggle to differentiate between genuine eigenstates and spurious invariant superpositions, potentially leading to convergence toward an incorrect solution.

Since we do not know  \textit{a priori} whether a fixed $\tau$ satisfies this problematic condition, a more robust strategy is to avoid using the same $\tau$ in every iteration. Instead, a different value of $\tau$ is randomly selected at each step from a suitably chosen interval. In this way, even if certain $\tau$ values prevent the algorithm from distinguishing some superpositions from true eigenstates, other values will not, thereby mitigating the occurrence of spurious eigenstates in the averaged behavior over multiple iterations.  For each pair of stationary states with distinct energies (i.e., $\omega_{\alpha,\beta} \neq 0$), the sinusoidal terms in Eq.~(\ref{survival_probability2}) have a period $2 \pi / \omega_{\alpha,\beta}$, so the longest period among them is $\tau_{\mathrm{max}} = 2 \pi / \omega_{\mathrm{min}}$, where $\omega_{\mathrm{min}}$ denotes the smallest nonzero $\omega_{\alpha,\beta}$. Sampling $\tau$ from an interval of width on the order of $\tau_{\mathrm{max}}$---for instance, $\tau \in [0, \tau_{\mathrm{max}}]$---helps to avoid systematic reinforcement of spurious solutions across iterations. If the spectrum of $H$ is unknown, as is the case in most practical situations, $\tau_{\mathrm{max}}$ can be roughly estimated from the characteristic energy scales of the problem.

 \subsection{Fine-tuning the algorithm after initial convergence}
 \label{subsec:algoft}
 A way to improve the precision of the algorithm is to introduce a fine-tuning stage once an initial convergence has been reached. Specifically, when the algorithm is considered to have converged---i.e., when the maximum exploration parameter satisfies $w_k^{(\mathrm{M})} < w_{\mathrm{th}}$---all exploration parameters $w_k^{(j,l)}$ are reset to a new value $w_{\mathrm{r}}$, chosen such that $w_{\mathrm{th}} < w_{\mathrm{r}} < 1$. This reinitialization introduces a new phase of exploration while preserving most of the knowledge already acquired.

The reset parameter $w_{\mathrm{r}}$ controls the balance between exploration and stability during the fine-tuning stage. If $w_{\mathrm{r}}$ is too large (close to $1$), the reset may partially disrupt the structure already learned by the algorithm. Conversely, if $w_{\mathrm{r}}$ is chosen too close to the threshold $w_{\mathrm{th}}$, the exploration becomes too limited to produce further improvements. A detailed discussion of the practical choice of $w_{\mathrm{r}}$, including parameter ranges for which the algorithm performs consistently, is provided in Sec.~\ref{subsec:hyperparameters}.

 If the reset procedure were applied indefinitely, the algorithm would never reach full convergence. To avoid this, the reset value $w_{\mathrm{r}}$ is gradually decreased after a fixed number of iterations. Specifically, once the iteration index exceeds a predetermined value $k_0$, $w_{\mathrm{r}}$ is reduced linearly according to the rule $w_{\mathrm{r}} ( k_{\mathrm{M}} - k ) / ( k_{\mathrm{M}} - k_0 )$,  where $k_{\mathrm{M}}$ denotes the maximum number of iterations allowed.
 
This schedule enables finer exploration after the initial convergence phase, giving the algorithm the flexibility to improve the solution while preserving the knowledge acquired in earlier stages.
 
\subsection{Assessing algorithm convergence and accuracy}
\label{subsec:algoaca}
Two key aspects must be considered when assessing the performance of the algorithm described above. First, the number of iterations required to achieve convergence should be as small as possible, since faster convergence reflects better algorithmic efficiency. Second, the algorithm must reproduce the target basis states $\mathsf{B}$ with high accuracy, ideally within the minimum number of iterations.

As discussed earlier, convergence can be quantified through the dependence of the maximum exploration parameter $w_k^{(\mathrm{M})}$ on the iteration index $k$. Regarding accuracy, in all the examples considered below the states of $\mathsf{B}$ can be computed independently, without relying on the algorithm. This makes it possible to evaluate accuracy directly by calculating the overlaps between the algorithmic states $\{ D_k \ket{j} \}_{j \in \{0,\dots,d-1\}}$ and the independently obtained vectors of $\mathsf{B}$. Since it is not known \textit{a priori} which vector of the target basis $\mathsf{B}$ a given computational basis state $\ket{j}$ will be mapped to,  it is necessary to adopt a flexible accuracy measure. A suitable measure of the algorithm’s accuracy is provided by the maximum square-root fidelities, defined for each $j \in \{0,\dots,d-1\}$ as
\begin{equation}
	\label{fidelities1}
	f_k^{(j)} = \max_{\alpha \in \{0,\dots,d-1\}} \abs{\mel{\Phi_{\alpha}}{D_k}{j}},
\end{equation}
which reflects the best overlap achieved by the transformed state $D_k\ket{j}$ with any state of the target basis $\mathsf{B}$ at iteration $k$.

It is important to note that the algorithm proposed here explores the space of unitary operators in a stochastic manner. This randomness is twofold: on one hand, the measurement outcomes in step 4 are intrinsically probabilistic; on the other hand, the angles $\alpha_k^{(j,l)}$, $\beta_k^{(j,l)}$, and $\gamma_k^{(j,l)}$ are generally chosen pseudo-randomly. Consequently, repeated applications of the algorithm generally produce different instances of $D_k$, each with varying accuracy. Even when the accuracy is high, in different realizations the same computational basis vector $\ket{j}$ may be mapped to different vectors in the target basis $\mathsf{B}$. Therefore, to properly assess the algorithm’s performance, it is convenient to consider a sufficiently large set of $N_{\mathrm{r}}$ realizations. For each iteration $k$, the arithmetic means of the maximum exploration parameter $w_k^{(\mathrm{M})}$ and the maximum square-root fidelity $f_k^{(j)}$ across realizations are computed and denoted as $W_k^{(\mathrm{M})}$ and $F_k^{(j)}$, respectively. These averaged quantities provide a robust way to quantify both the convergence and accuracy of the algorithm over multiple stochastic realizations.

In what follows, we test the convergence and accuracy of the proposed algorithm by applying it to the calculation of the eigenstates of several representative Hamiltonian models. For this purpose, all Hamiltonians are rescaled to a dimensionless form,
\begin{equation}
	\label{rescaleH}
	\tilde{H} = \frac{H - E_{\rm min} I}{E_{\rm max} - E_{\rm min}},
\end{equation}
where $E_{\rm min}$ and $E_{\rm max}$ are the minimum and maximum eigenenergies of $H$, respectively. This guarantees that the spectrum of $\tilde{H}$ lies between $0$ and $1$, with the lowest eigenvalue equal to $0$ and the highest equal to $1$. In addition, we define a dimensionless evolution time as $\tilde{\tau} = \tau (E_{\rm max} - E_{\rm min})/\hbar$, so that the time parameter naturally adapts to the spectral range of the Hamiltonian. To avoid convergence of the algorithm to spurious eigenstates, instead of keeping $\tau$ (or its dimensionless version $\tilde{\tau}$) fixed, at each iteration we randomly select its value within a suitable interval, as described in Sec.~\ref{subsec:algofpa}.

\subsection{Testing the algorithm on random Hamiltonians}
\label{subsec:algorH}
To provide an initial illustration of the proposed algorithm, we apply it to the calculation of the eigenvectors of random Hamiltonians. In each realization, a different Hermitian matrix with randomly chosen complex entries is generated, and the associated Hamiltonian is rescaled according to Eq.~(\ref{rescaleH}). The algorithm is then executed for a sufficient number of iterations to ensure convergence, and the obtained states are compared with the exact eigenvectors computed by direct diagonalization, using the fidelities defined in Eq.~(\ref{fidelities1}). Repeating this procedure over $N_{\mathrm{r}}$ realizations finally allows us to compute the mean fidelities $F_k^{(j)}$ and the mean maximum exploration parameter $W_k^{(\mathrm{M})}$.

\begin{figure}[htbp]
	\includegraphics[width=0.48\textwidth]{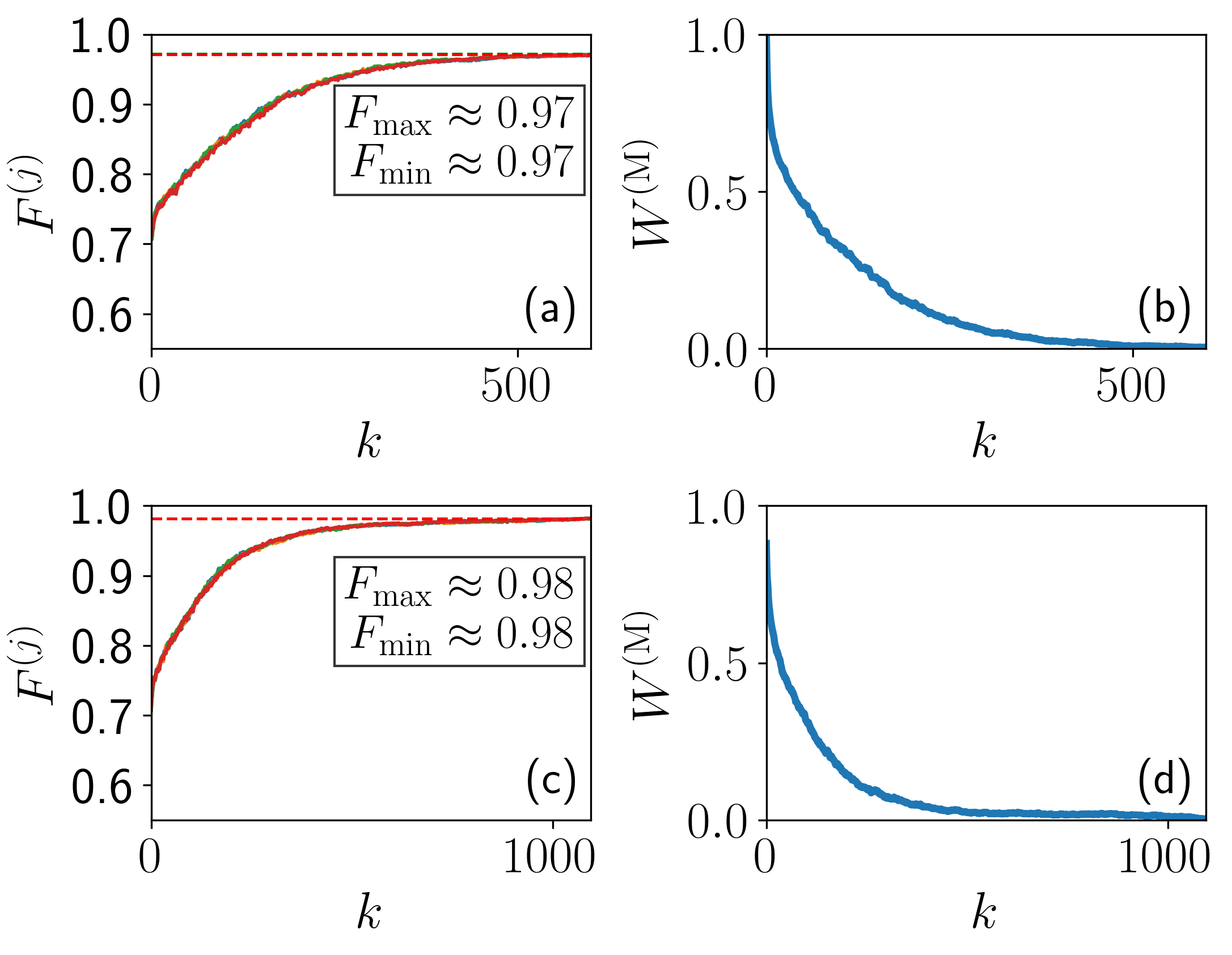}
	\caption{Algorithm results for random two-qubit Hamiltonians ($d=4$). Panels (a) and (b) show the fidelities $F_k^{(j)}$ and the maximum exploration parameter $W_k^{(\mathrm{M})}$ as functions of the iteration number $k$ without the reset mechanism described in Sec.~\ref{subsec:algoft}, while panels (c) and (d) show the results with reset. Each realization uses a different random Hamiltonian, and in each iteration the dimensionless evolution time $\tilde{\tau}$ is uniformly sampled from $[0,100]$. Parameters are $r=0.9$, $p=2/r$, $w_{\mathrm{th}}=0.005$, $w_{\mathrm{r}}=0.01$, and $N_{\mathrm{r}}=1000$. Fidelities for the four computational basis states are plotted in different colors; however, because the curves largely overlap, the colors are difficult to distinguish.  In the left panels, dashed horizontal lines indicate the maximum and minimum fidelities, $F_{\mathrm{max}}$ and $F_{\mathrm{min}}$, defined as the maximum and minimum values of $F_k^{(j)}$ over all computational basis states $\ket{j}$ in the converged regime (i.e., for sufficiently large iteration number $k$). The near equality of $F_{\mathrm{max}}$ and $F_{\mathrm{min}}$ reflects the uniformity of the learning process across basis states.}
	\label{fig2}
\end{figure}

Figure~\ref{fig2} shows the results obtained for a two-qubit system ($d=4$). Panels (a) and (b) display the behavior of $F_k^{(j)}$ and $W_k^{(\mathrm{M})}$ as a function of the iteration number $k$ without the reset mechanism described in Sec.~\ref{subsec:algoft}, while panels (c) and (d) show the corresponding results when the reset is applied. In both cases, in each iteration the dimensionless evolution time $\tilde{\tau}$ was uniformly sampled from the interval $[0,100]$ to avoid the issue discussed in Sec.~\ref{subsec:algofpa}. The remaining parameters are set to $r=0.9$, $p=2/r$, $w_{\mathrm{th}}=0.005$, $w_{\mathrm{r}}=0.01$, and  $N_{\mathrm{r}}=1000$. The fidelities corresponding to the four possible states of the computational basis are represented in different colors. However, since the curves largely overlap, the colors cannot be easily distinguished. This strong overlap reflects the fact that the learning dynamics is essentially identical for all basis states, indicating that both the accuracy and the convergence rate are uniform across the computational basis. This behavior is not generic, as will be illustrated in the pairing model discussed in Sec.~\ref{subsec:pairing}, where the convergence depends on the symmetry sector.

As shown in Fig.~\ref{fig2}, the algorithm attains fidelities close to unity both with and without the reset mechanism, with a slight improvement when resets are applied. The convergence of the algorithm is characterized by the behavior of the exploration parameter $W_k^{(\mathrm{M})}$, which saturates at small values for large iteration number $k$. The average number of iterations required to reach this regime is lower in the absence of resets. This can be attributed to the fact that the reset scheme introduces additional exploration stages, which increases the total number of iterations needed to reach convergence. To quantify the accuracy across the computational basis, we define $F_{\mathrm{max}}$ and $F_{\mathrm{min}}$ as the maximum and minimum values of $F_k^{(j)}$ over all basis states $\ket{j}$ in the converged regime (i.e., for sufficiently large $k$). In this regime, these quantities are nearly identical: without resets $F_{\mathrm{max}} \approx F_{\mathrm{min}} \approx 0.97$, whereas with resets $F_{\mathrm{max}} \approx F_{\mathrm{min}} \approx 0.98$. The near equality $F_{\mathrm{max}} \approx F_{\mathrm{min}}$ provides a direct measure of the uniformity of the learning process across basis states.

\begin{figure}[htbp]
	\includegraphics[width=0.48\textwidth]{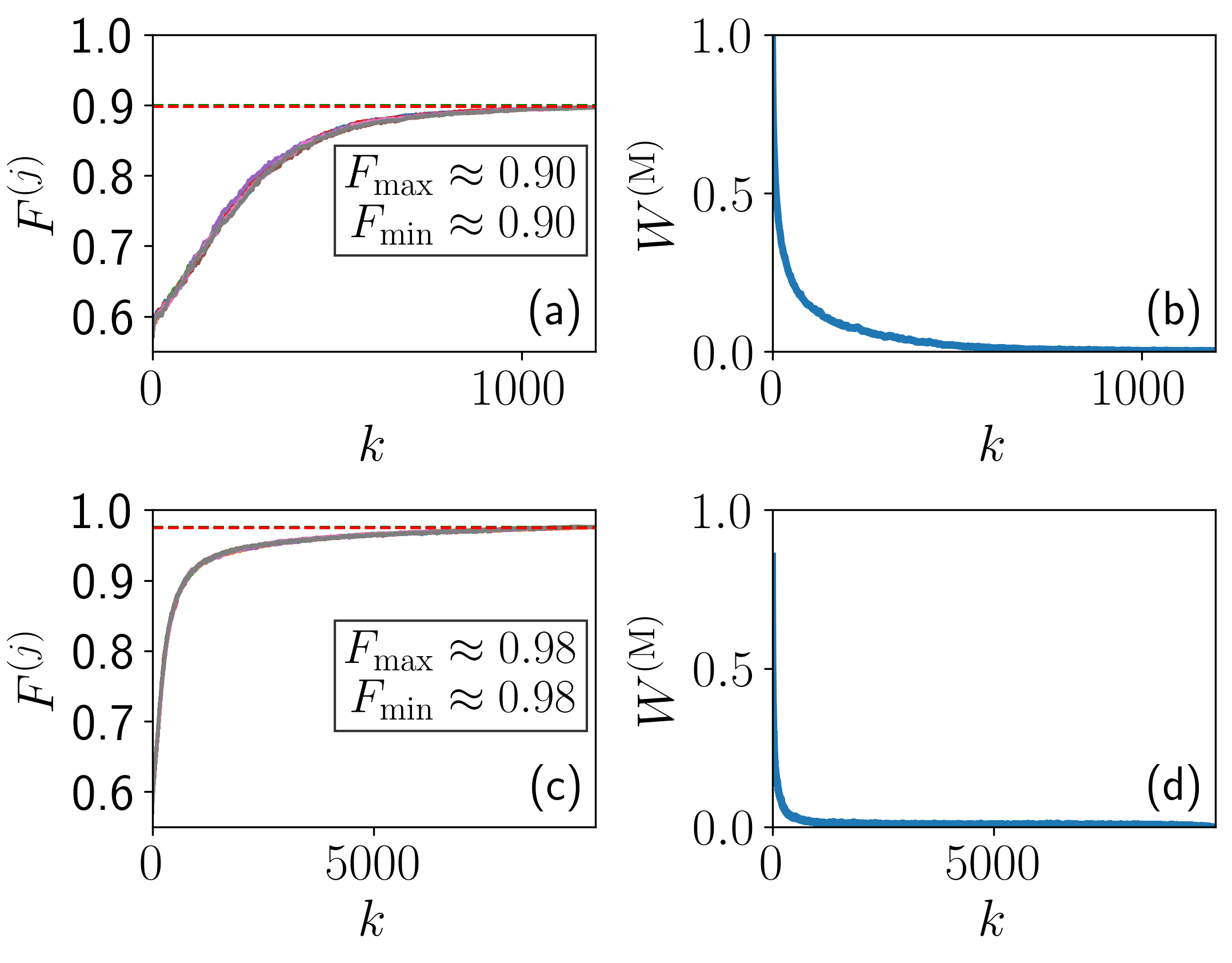}
	\caption{Same as Fig.~\ref{fig2}, but for the case of three qubits ($d=8$). All parameters are identical to those used for the two-qubit case shown in Fig.~\ref{fig2}, except for the increased Hilbert-space dimension.}
	\label{fig3}
\end{figure}

In Fig.~\ref{fig3}, we present results analogous to those in Fig.~\ref{fig2}, but for a three-qubit system ($d=8$). The reset mechanism described in Sec.~\ref{subsec:algoft} substantially improves the fidelities, increasing them from $F_{\mathrm{min}} \approx F_{\mathrm{max}} \approx 0.90$ without resets to  $F_{\mathrm{min}} \approx F_{\mathrm{max}} \approx 0.98$ with resets. While this improvement comes with an increase in the number of iterations required for convergence, the reset scheme consistently guides the algorithm toward high-quality results across all basis states, demonstrating its effectiveness in larger systems.

Random Hamiltonians of the type considered in this section have also been studied in the literature  using a reinforcement learning algorithm different from the one presented here (see Ref.~\cite{Albarran_Arriagada_2020}). Providing a detailed description of the protocol of Ref.~\cite{Albarran_Arriagada_2020} is beyond the scope of this work. Nevertheless, although both approaches use measurement outcomes as the basis for reward and punishment, there are fundamental differences between the two reinforcement learning strategies. In particular, while the method of Ref.~\cite{Albarran_Arriagada_2020} searches for eigenstates sequentially, starting from an arbitrarily chosen one, our approach is designed to learn all eigenstates simultaneously.

When comparing the results presented here with those of Ref.~\cite{Albarran_Arriagada_2020} for the two-qubit case---the only one reported in that work---we observe that our algorithm achieves slightly higher fidelities across the full set of states. In the method of Ref.~\cite{Albarran_Arriagada_2020}, some states are obtained with high accuracy, whereas others perform less favorably. This difference may be related to the fact that, in our approach, all states are treated on equal footing, while in the other method some states are obtained earlier in the sequential procedure. Another important aspect is that the number of iterations required for convergence is significantly smaller in our algorithm. In fact, when attempting to extend the protocol of Ref.~\cite{Albarran_Arriagada_2020} to three qubits, we found that convergence could be reached for some states, but obtaining convergence for all of them within a reasonable number of iterations proved challenging. In the following subsection, we discuss the selection of the hyperparameters and the robustness of the algorithm with respect to their variation. We then apply the algorithm in Sec.~\ref{sec:applications} to paradigmatic physical Hamiltonians, including the transverse-field Ising model (TFIM) and the Richardson pairing Hamiltonian, to demonstrate its applicability and performance beyond random systems.

\subsection{Hyperparameter selection and robustness}
\label{subsec:hyperparameters}

The reinforcement-learning protocol involves several hyperparameters controlling both the learning dynamics and its practical implementation. These include the reward and punishment rates $r$ and $p$, the convergence threshold $w_{\mathrm{th}}$, the time scale $\tau_{\max}$ defining the range of sampled evolution times, and the reset parameter $w_{\mathrm{r}}$. Here, we provide a discussion of their practical selection and impact on the algorithmic performance.

We first discuss the role of the reward and punishment rates. Our numerical results indicate that choosing the product $rp = 2$ provides consistently good performance across all the models considered in this work. Within this constraint, the choice of $r$ determines the balance between exploration and exploitation during the learning process. These numerical results suggest a simple practical guideline: fixing $rp = 2$ and selecting $r$ as large as possible within the convergence regime provides reliable performance across different models.

\begin{figure}[htbp]
	\includegraphics[width=0.48\textwidth]{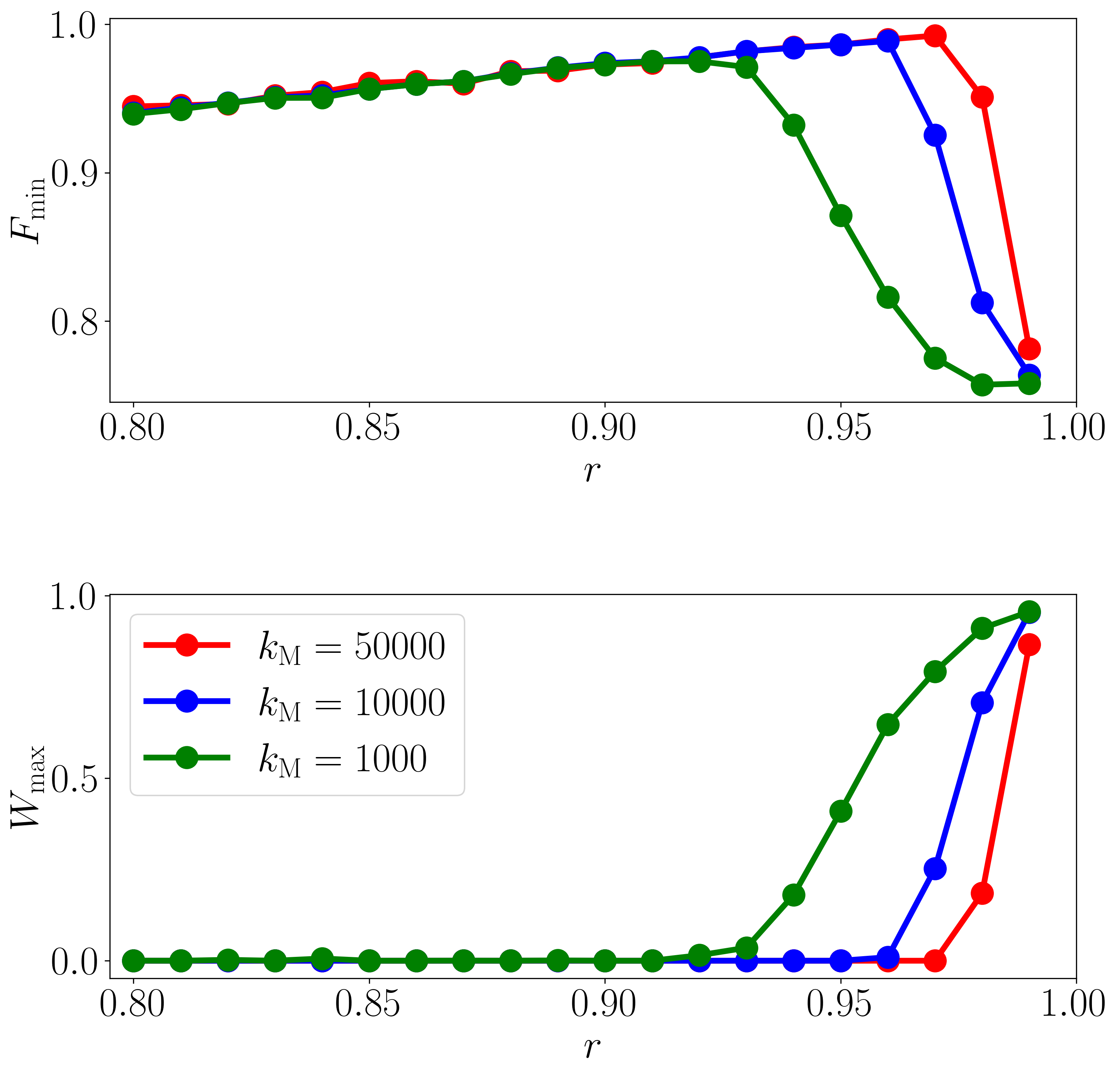}
\caption{Dependence of the algorithmic performance on the reward parameter $r$ for a two-qubit system with a random Hamiltonian, in the absence of reset, and with fixed $rp=2$. The evolution time $\tilde{\tau}$ is sampled uniformly in the interval $[0,100]$, and the convergence threshold is set to $w_{\mathrm{th}}=0.005$. Results are averaged over  $N_{\mathrm{r}}=100$ realizations.
Top panel: minimum fidelity over all learned states at the end of the protocol, $F_{\mathrm{min}}$, providing a lower bound on the fidelities of all states. Bottom panel: maximum value of $W_k^{(\mathrm{M})}$ at the final iteration, $W_{\mathrm{max}}$, providing an upper bound on the convergence of the algorithm.
Different colors correspond to different values of the maximum number of iterations: $k_{\mathrm{M}}=10^3$ (green), $10^4$ (blue), and $5\times 10^4$ (red).}
	\label{fig_new}
\end{figure}

To illustrate the effect of $r$ under the constraint $rp = 2$, we consider a two-qubit system with a random Hamiltonian in the absence of reset. Figure~\ref{fig_new} shows the dependence of the algorithmic performance on $r$ for different values of the maximum number of iterations $k_{\mathrm{M}}$. The upper panel displays the minimum fidelity over all learned states at the end of the protocol, $F_{\mathrm{min}}$, while the lower panel shows the maximum value of $W_k^{(\mathrm{M})}$ at the final iteration, $W_{\mathrm{max}}$.

As shown in the figure, increasing the number of iterations extends the range of values of $r$ for which the algorithm converges, allowing values closer to $r=1$ to be used. Within the convergence regime, larger values of $r$ lead to improved fidelities, as reflected by higher values of $F_{\mathrm{min}}$. For values of $r$ beyond this regime, the algorithm does not converge within the allotted number of iterations, resulting in a degradation of the fidelities. These results indicate that, for a given iteration budget, a suitable choice of $r$ corresponds to the largest value compatible with convergence.

The convergence threshold $w_{\mathrm{th}}$ defines the numerical criterion used to determine when convergence is reached and does not affect the learning dynamics. In all simulations presented in this work, we use a fixed value $w_{\mathrm{th}} = 0.005$.

The parameter $\tau_{\max}$ defines the range from which the evolution time $\tau$ is sampled during the protocol. As discussed in Sec.~\ref{subsec:algofpa}, its choice is related to the characteristic energy scales of the problem.
For instance, for the rescaled Hamiltonian in Eq.~(\ref{rescaleH}), the dimensionless energy spectrum spans an interval of width one. Assuming approximately equally spaced energy levels, the typical level spacing can be estimated as $1/(d-1)$. On this basis, it is sufficient to choose the upper bound of the sampling interval for the dimensionless parameter $\tilde{\tau}$ large compared to $2\pi (d-1)$ in order to suppress the reinforcement of spurious fixed points. Once this condition is satisfied, the precise choice of this upper bound has only a moderate impact on the achieved fidelities and on the convergence behavior of the algorithm.

Concerning the reset parameter $w_{\mathrm{r}}$, we find that the algorithm performs consistently for values of $w_{\mathrm{r}}$ in the interval $[0.01,0.05]$ across all the models considered in this work, suggesting that this range is not strongly problem-dependent. Within this range, variations of $w_{\mathrm{r}}$ lead only to moderate quantitative changes in the achieved fidelities and convergence rates, without affecting the qualitative behavior of the algorithm.

Taken together, these observations indicate that the performance of the protocol is robust with respect to variations of the hyperparameters within the ranges discussed above. In particular, the method operates reliably within broad parameter ranges and does not require precise fine tuning to achieve consistent performance across different Hamiltonians.

\section{Application to physical systems}
\label{sec:applications}

In this section, we apply the method proposed above to two types of physical systems. The first, relevant to condensed matter physics, is the TFIM~\cite{Fradking2013}, which is commonly used to describe condensed matter systems and has also frequently served as a benchmark in quantum computing studies. Assuming only nearest-neighbor interactions, the TFIM Hamiltonian can be written as~\cite{Anselme2024}
\begin{equation}
H = -J\sum_{j=0}^{N-2} Z_j Z_{j+1} - h\sum_{j=0}^{N-1} X_j- K \sum_{j=0}^{N-2} X_j Z_{j+1}.
 \label{eq:HTFIM}
\end{equation}
Here $j\in\{0, \cdots , N-1\}$ labels the $N$ qubits, $\{X_j, Y_j, Z_j\}$ denotes the set of associated Pauli matrices, and $J$, $h$, and $K$ are parameters with dimensions of energy.

The second model is the pairing Hamiltonian, also known as the Richardson Hamiltonian~\cite{Dukelsky2004,VONDELFT2001,Zelevinsky2003,Brink2005}, which is often used to describe small superfluid systems, such as atomic nuclei. Here, we focus on the specific subspace with seniority zero (no pair breaking) and employ the pair-to-qubit encoding technique. In this encoding, the Hamiltonian reads~\cite{Ruiz2022}
 \begin{equation}
	H = \sum_{j=0}^{N-1} \!\left(\varepsilon_{j}-\frac{g}{2}\right) \!\left(I - Z_j\right)
		- \frac{g}{2} \sum_{j=1}^{N-1}\sum_{k=0}^{j-1}\left( X_j X_k + Y_j Y_k \right),
\label{eq:Hpairencoding}
\end{equation}
where $\varepsilon_{i}$ and $g$ are parameters with dimensions of energy, representing the single-particle energy levels and the interaction strength, respectively. In the following, we assume equidistant single-particle levels, $\varepsilon_j = j \Delta \varepsilon$, with $\Delta \varepsilon$ denoting the energy spacing between levels.

Both models share the feature of exhibiting a quantum phase transition (QPT). In the TFIM, the transition occurs from ordered to disordered spin orientations, whereas in the pairing model it corresponds to a $U(1)$ symmetry-breaking transition associated with particle-number conservation. Beyond the presence of QPTs, however, the two models represent very different physical systems, both in nature and in their quantum information content. Specifically, the TFIM involves only local interactions and is generally characterized by area-law entanglement, while the pairing model features all-to-all interactions and is typically associated with volume-law entanglement.

In the following subsections, we present the results obtained by implementing the algorithm on these two models. As in the case of the random Hamiltonians analyzed earlier, both the TFIM and pairing Hamiltonians will first be rescaled to dimensionless form according to Eq.~(\ref{rescaleH}). In addition, at each iteration the dimensionless time $\tilde{\tau}$ will be randomly sampled within an appropriate interval, as described in Sec.~\ref{subsec:algofpa}, to prevent convergence to spurious eigenstates.

\subsection{Application to the TFIM model}

\begin{figure}[t]
	\includegraphics[width=0.48\textwidth]{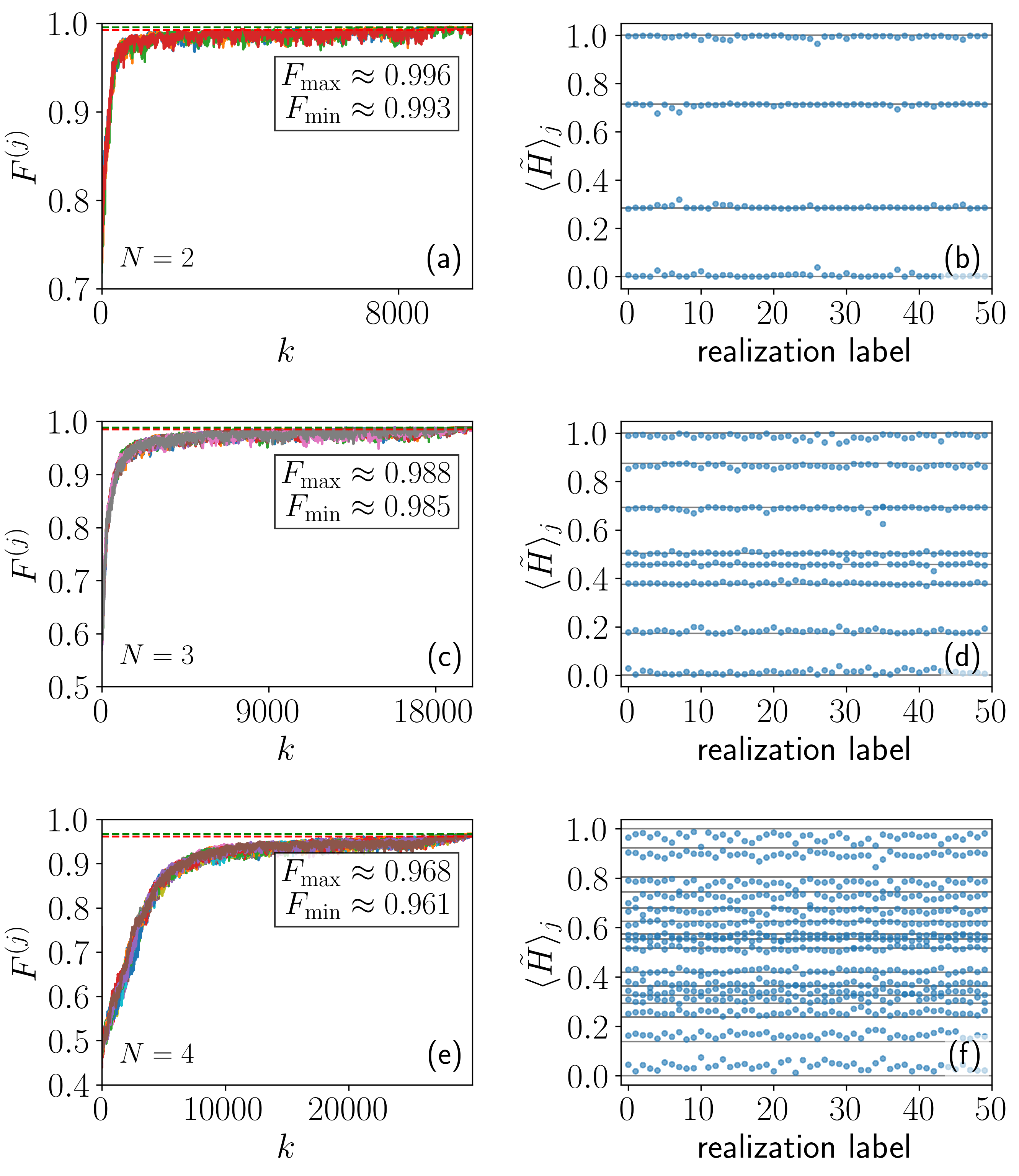}
	\caption{Results of the algorithm applied to the TFIM Hamiltonian in Eq.~(\ref{eq:HTFIM}) with $J/h = 1$ and $K/h = 0.5$. The remaining parameters are $r = 0.9$, $p = 2/r$, $w_{\mathrm{th}} = 0.005$, $w_{\mathrm{r}} = 0.05$, and $N_{\mathrm{r}} = 50$. In each iteration, the dimensionless evolution time $\tilde{\tau}$ is uniformly sampled from $[0,600]$. The upper panels [(a) and (b)] correspond to the two-qubit case ($d=4$), the middle panels [(c) and (d)] to three qubits ($d=8$), and the lower panels [(e) and (f)] to four qubits ($d=16$). In the left panels [(a),(c), and (e)], the mean fidelities $F_k^{(j)}$ associated with each computational-basis state $\ket{j}$ are shown as a function of the iteration number $k$, with different colors indicating different states. Since many curves overlap, color differences may be difficult to distinguish.  Dashed horizontal lines indicate the maximum and minimum fidelities, $F_{\mathrm{max}}$ and $F_{\mathrm{min}}$.  The right panels [(b), (d), and (f)] display, for each computational basis state $\ket{j}$, the dimensionless expectation values of the energies obtained after each stochastic realization has converged, plotted against the corresponding realization index. In these panels, horizontal solid lines represent the exact eigenenergies computed by numerical diagonalization. }
	\label{fig4}
\end{figure}

To evaluate the efficiency of the algorithm when applied to the TFIM Hamiltonian, we first analyze the mean fidelities $F_k^{(j)}$ shown in the left panels of Fig.~\ref{fig4}. In these simulations, following the procedure described in Sec.~\ref{subsec:algofpa}, the dimensionless propagation time $\tilde{\tau}$ is randomly sampled at each iteration within the interval $[0,600]$ to prevent convergence to spurious stationary states. In addition, the reset mechanism outlined in Sec.~\ref{subsec:algoft} is applied with parameters $w_{\mathrm{th}} = 0.005$ and  $w_{\mathrm{r}}=0.05$. As shown in the figure, a slight decrease in both the maximum and minimum fidelities is observed as the number of qubits increases: $F_{\mathrm{max}}=0.996$ and $F_{\mathrm{min}}=0.993$ for two qubits, $F_{\mathrm{max}}=0.988$ and $F_{\mathrm{min}}=0.985$ for three qubits, and $F_{\mathrm{max}}=0.968$ and $F_{\mathrm{min}}=0.961$ for four qubits. This reduction in fidelities is expected, as increasing the number of qubits exponentially enlarges the dimension of the Hilbert space, making the exploration of the state space more challenging.

In the right panels of Fig.~\ref{fig4}, we show, for each stochastic realization and for each computational basis state $\ket{j}$, the  expectation values of the dimensionless TFIM Hamiltonian after the algorithm has converged, plotted against the index assigned to each realization. Specifically, for sufficiently large values of $k$ such that the algorithm has effectively converged, we represent 
\begin{equation}
\langle \tilde{H}\rangle_j = \expval{D_k^\dagger \tilde{H} D_k }{j}
\label{expvaldef}
\end{equation}
 for each $j \in \{0, \dots, d-1\}$ and for each realization, with $\tilde{H}$ given by Eq.~(\ref{rescaleH}) using the TFIM Hamiltonian of Eq.~(\ref{eq:HTFIM}). The solid horizontal lines indicate the exact values obtained via numerical diagonalization of the Hamiltonian. As can be seen, for the two- and three-qubit cases, the energies obtained in most realizations match the exact numerical values quite closely. This indicates that, in these cases, even a single realization is sufficient to approximate the full set of eigenstates. For four qubits, however, deviations become more noticeable, even though the fidelities are only slightly lower than in the smaller systems. This highlights that the computation of these energies is highly sensitive to small variations in the stationary states approximated by the algorithm.

\begin{figure}[t]
	\includegraphics[width=0.48\textwidth]{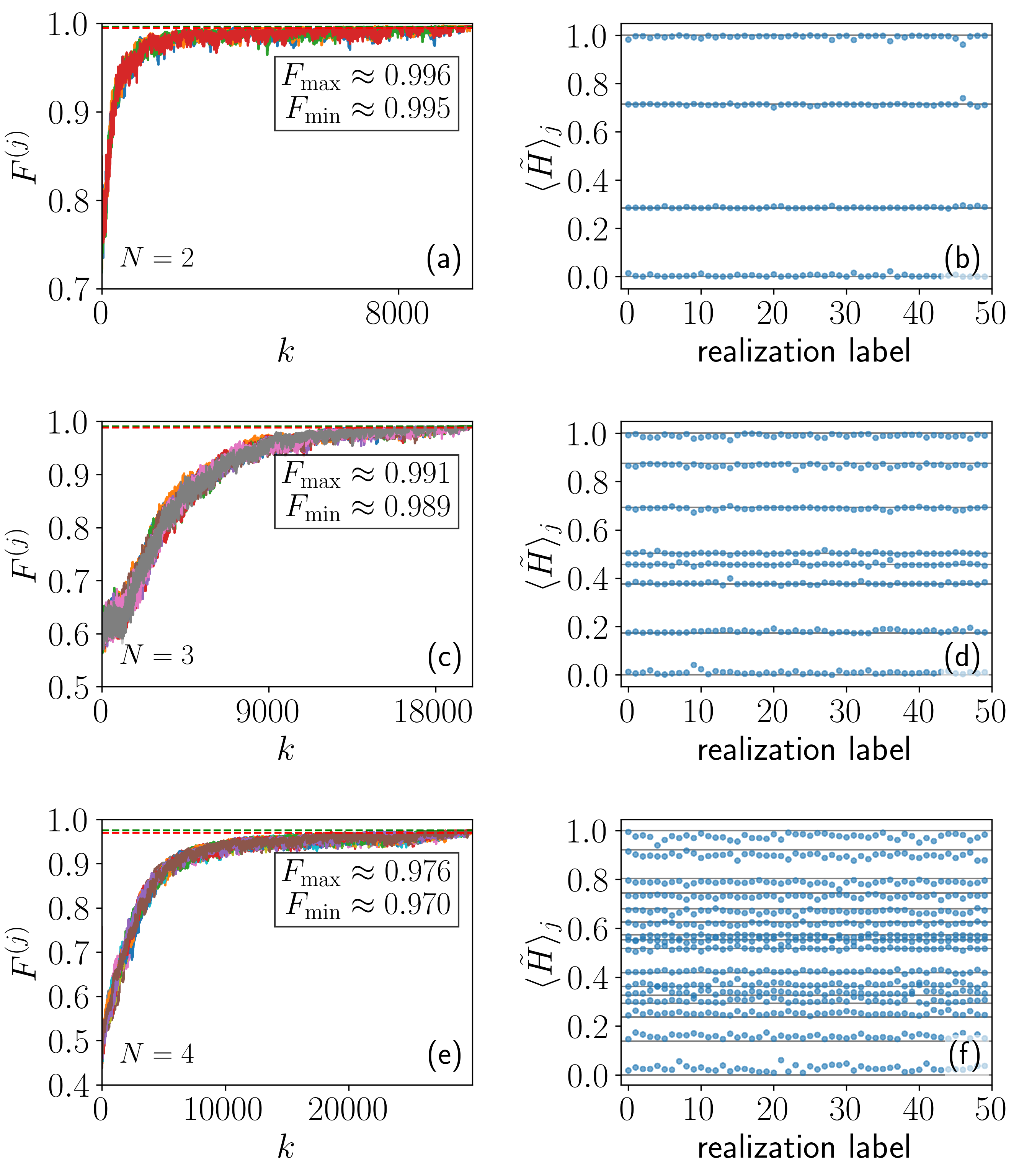}
	\caption{Results analogous to those shown in Fig.~\ref{fig4}, obtained for $r = 0.93$. All other parameters are identical to those used in Fig.~\ref{fig4}.}
	\label{fig5}
\end{figure}

In an effort to improve the estimation of the eigenenergies, we slightly modified the algorithm parameters. Specifically, in Fig.~\ref{fig5} we show the same results as in Fig.~\ref{fig4}, but using a slightly higher reward parameter, $r = 0.93$. Compared to Fig.~\ref{fig4}, this adjustment of the reward parameter generally leads to a modest improvement in both the fidelities and the estimated energies, indicating that small changes in the algorithm settings can enhance the accuracy of the results across all realizations.

It should be noted that the example discussed above represents an ideal benchmark, as the performance of the algorithm can be directly assessed by comparing its output with the exact stationary states and eigenenergies obtained through numerical diagonalization of the Hamiltonian. In more realistic scenarios, however, such exact information may not be available, making it impossible to evaluate the accuracy of the algorithm in a direct manner. Nevertheless, for the eigenenergies, it is still possible to identify, among all stochastic realizations of the algorithm, those providing the best approximations to the ground and highest-energy states. Specifically, the realizations yielding the smallest and largest expectation values of the Hamiltonian can be associated with these two states, respectively. This identification is justified by the variational principle, which states that the expectation value of the Hamiltonian in any normalized state constitutes an upper bound to the exact ground-state energy. Likewise, applying the same argument to the inverted Hamiltonian, $-H$, the expectation value provides a lower bound to the highest-energy eigenvalue. Consequently, the realizations corresponding to the smallest and largest energy expectations offer the most accurate approximations to the ground and highest-energy states, respectively. Later, we will introduce a more general criterion based on evaluating the energy dispersions of the states obtained from the different realizations of the algorithm.

\subsection{Application to the pairing model}
\label{subsec:pairing}

We now consider the pairing Hamiltonian given by Eq.~(\ref{eq:Hpairencoding}). This Hamiltonian describes a set of interacting particle pairs and is relevant for modeling the transition from normal to superfluid phases in small superconducting systems. Notably, while in the TFIM model the exact solution required the diagonalization of a $d \times d$ matrix, here the Hamiltonian exhibits a block-diagonal structure in the computational qubit basis due to particle number conservation. Specifically, after applying the particle-to-qubit transformation leading to Eq.~(\ref{eq:Hpairencoding}), particle number conservation is mapped onto the conservation of the Hamming weight. Starting from a given basis state $\ket{j}$, we can equivalently write it in the qubit basis as $\ket{[j]}$, where $[j]$ denotes the binary representation of $j$. The Hamming weight is defined as the number of ``1''s in $[j]$. In the pairing Hamiltonian, two basis states $\ket{j}$ and $\ket{k}$ have non-zero Hamiltonian matrix elements only if they share the same Hamming weight. Consequently, the block corresponding to Hamming weight $N_{\mathrm{Ham}}$ has dimension $C^{N_{\mathrm{Ham}}}_N$, the binomial coefficient (``$N$ choose $N_{\mathrm{Ham}}$''), which counts the number of ways to choose $N_{\mathrm{Ham}}$ objects out of $N$. This block structure is commonly exploited in classical computations to access eigenstates by diagonalizing the Hamiltonian within each block. In the context of machine learning, however, exploiting problem symmetries can introduce additional challenges. Here, one of our objectives is to test whether the RL method can converge to the correct energies without explicitly enforcing these symmetries during learning.

\begin{figure}[t]
	\includegraphics[width=0.48\textwidth]{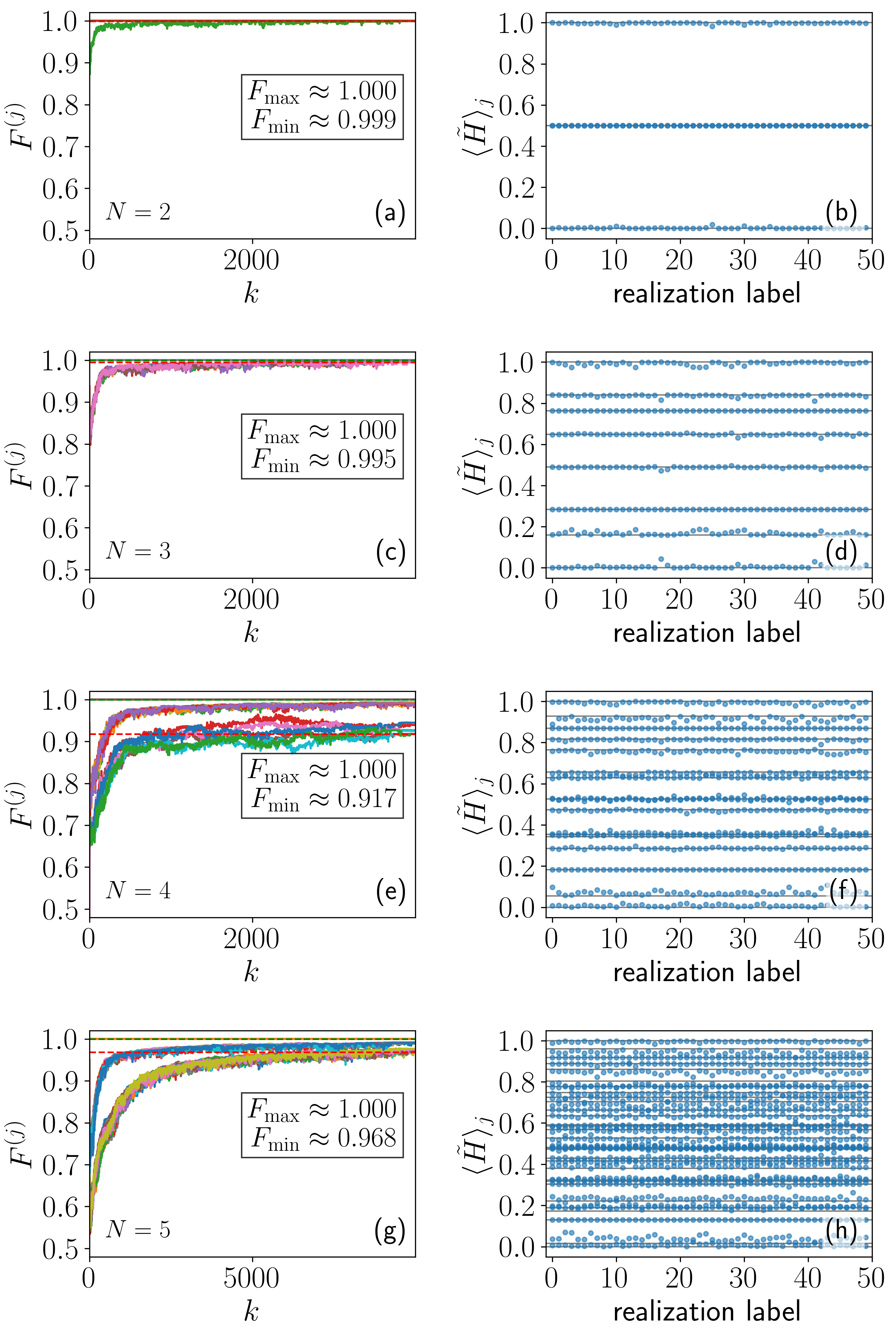}
	\caption{Results of the algorithm applied to the pairing Hamiltonian in Eq.~(\ref{eq:Hpairencoding}) with $\varepsilon_j = j \Delta \varepsilon$ and $g / \Delta \varepsilon = 1$. The remaining parameters are $r = 0.9$, $p = 2/r$, $w_{\mathrm{th}} = 0.005$, $w_{\mathrm{r}} = 0.05$, and $N_{\mathrm{r}} = 50$. In each iteration, the dimensionless evolution time $\tilde{\tau}$ is uniformly sampled from $[0,600]$. Panels (a)–(b) show the two-qubit case ($d = 4$), (c)–(d) the three-qubit case ($d = 8$), (e)–(f) the four-qubit case ($d = 16$), and (g)–(h) the five-qubit case ($d = 32$). The left panels show the mean fidelities $F_k^{(j)}$ associated with each computational-basis state $\ket{j}$ as a function of the iteration number $k$, with different colors indicating different states. Since many curves overlap, color differences may be difficult to distinguish. Dashed horizontal lines indicate the maximum and minimum fidelities, $F_{\mathrm{max}}$ and $F_{\mathrm{min}}$. The right panels display, for each computational-basis state $\ket{j}$, the dimensionless expectation values of the energies obtained after each stochastic realization has converged, plotted against the corresponding realization index. In these panels, horizontal solid lines represent the exact eigenenergies computed by numerical diagonalization.}
	\label{fig6}
\end{figure}

Figure~\ref{fig6} presents results analogous to those in Fig.~\ref{fig4}, but for the pairing Hamiltonian instead of the TFIM model. The left panels display the mean fidelities for each computational-basis state $\ket{j}$, averaged over all stochastic realizations, as a function of the iteration number. The right panels show, for each realization and computational-basis state, the expectation values of the dimensionless pairing Hamiltonian obtained after the algorithm has converged, compared with the exact eigenenergies from numerical diagonalization. The results are shown for increasing numbers of qubits, from $N = 2$ to $N = 5$. The coupling strength parameter is set to $g / \Delta \varepsilon = 1$, corresponding to the strong-coupling regime that gives rise to the superfluid phase. Accordingly, the eigenstates generally involve complex superpositions of the original basis states. However, there are two particular computational basis states, $\ket{0}$ and $\ket{d-1}$, which correspond to Hamming weights $N_{\mathrm{Ham}} = 0$ and $N_{\mathrm{Ham}} = N$, respectively, and are exact eigenstates of the pairing Hamiltonian. Consequently, two horizontal lines with fidelity equal to $1$ appear in the left panels of the figure, and the maximum fidelity is therefore always $1$ in those plots.

Although the properties of the pairing Hamiltonian differ substantially from those of the TFIM model, we again observe that, after a reasonable number of iterations, the algorithm achieves very good fidelities in most cases, with the exception of the four-qubit system. As seen in the left panels, the mean fidelities remain high overall, although the minimum fidelity tends to decrease as the number of qubits increases. This trend is expected, since the dimension of the Hilbert space grows exponentially with the system size, making the exploration of the state space increasingly demanding. This behavior is consistent with the results shown in the right panels, where the expectation values of the dimensionless Hamiltonian are generally very close to the exact eigenenergies across nearly all realizations. An exception occurs for $N = 4$, where two distinct groups of fidelities appear in panel (e): one set converging close to unity and another slightly above $0.9$, resulting in a minimum fidelity smaller than that for $N = 5$. This apparent inconsistency with the accurate reproduction of the eigenenergies [panel (f)] can be explained by the presence of an exact degeneracy in one of the eigenenergies. Within the corresponding degenerate subspace, the eigenstates are not uniquely defined, and the algorithm may converge to different linear combinations than those obtained from direct diagonalization, leading to slightly reduced fidelities. Nevertheless, the method performs well for $N = 4$, successfully recovering the correct set of eigenenergies.

For $N = 5$ [panels (g) and (h)], three distinct groups of curves appear, all reaching high fidelities but requiring different numbers of iterations to converge. In the final iterations, the expectation values of the dimensionless Hamiltonian remain reasonably close to the exact energies, though the comparison becomes less straightforward due to the increased density of states.

The emergence of multiple convergence rates arises from the symmetry structure of the pairing Hamiltonian, which organizes the Hilbert space into blocks of different sizes according to particle-number conservation. In the qubit representation, these blocks correspond to subspaces with fixed Hamming weight. For five qubits, there are two blocks of size 1 ($N_{\mathrm{Ham}} = 0$ and $N_{\mathrm{Ham}} = 5$), two of size 5 ($N_{\mathrm{Ham}} = 1$ and $N_{\mathrm{Ham}} = 4$), and two of size 10 ($N_{\mathrm{Ham}} = 2$ and $N_{\mathrm{Ham}} = 3$). The cases $N_{\mathrm{Ham}} = 0$ and $N_{\mathrm{Ham}} = 5$ correspond to the exact eigenstates $\ket{0}$ and $\ket{31}$, explaining the horizontal lines with fidelity one in panel (g). The groups of curves that converge faster correspond to the blocks of size 5, while the slower convergence is associated with the larger blocks of size 10. This behavior reflects that the number of iterations required for convergence increases with the dimension of the explored subspace. It is worth noting that the algorithm manages to converge within all symmetry sectors without being provided with any prior information about the underlying symmetries.

\subsection{Symmetry-restricted RL}
\label{subsec:sym_rest_RL}

We showed in the previous section that the RL protocol introduced here is able to capture the symmetries of a Hamiltonian without prior knowledge of them. Alternatively, if the symmetries of the system are known in advance, the RL protocol can be directly applied within a given symmetry sector. Exploiting such symmetries can effectively extend the applicability of the RL technique to systems with a larger number of qubits, since each sector has a lower dimension than the full Hilbert space.

To illustrate the above-mentioned procedure, we first consider the case $N = 5$ shown in panels (g) and (h) of Fig.~\ref{fig6}. To account for particle-number conservation, instead of performing the RL in the full Hilbert space of dimension $d = 2^5 = 32$, we apply the RL algorithm separately within sectors of states $\ket{j}$ whose Hamming weight takes a fixed value between $0$ and $5$. 

To implement the restriction of the RL protocol to a fixed Hamming-weight sector, we use the same algorithm as in the general case, but restrict its action to the subset of basis states with a given Hamming weight $N_{\rm Ham}$.
			In practice, we identify the active subspace by iterating over the computational basis states labeled by $j$ and selecting those whose binary representation contains exactly $N_{\rm Ham}$ ones. States that do not satisfy this condition are treated as inactive.
		Inactive states are excluded from all stages of the algorithm: they are neither evolved nor measured, and they are not included in the set of pairs $(j,l)$ used during the reward/punishment updates (see item~5 in Sec.~\ref{subsec:algostep} and Appendix~\ref{appendix}). As a result, the learning dynamics is effectively restricted to the chosen symmetry sector. Notably, the RL protocols applied in different symmetry sectors are independent of each other.
        
\begin{figure}[t]
	\includegraphics[width=0.48\textwidth]{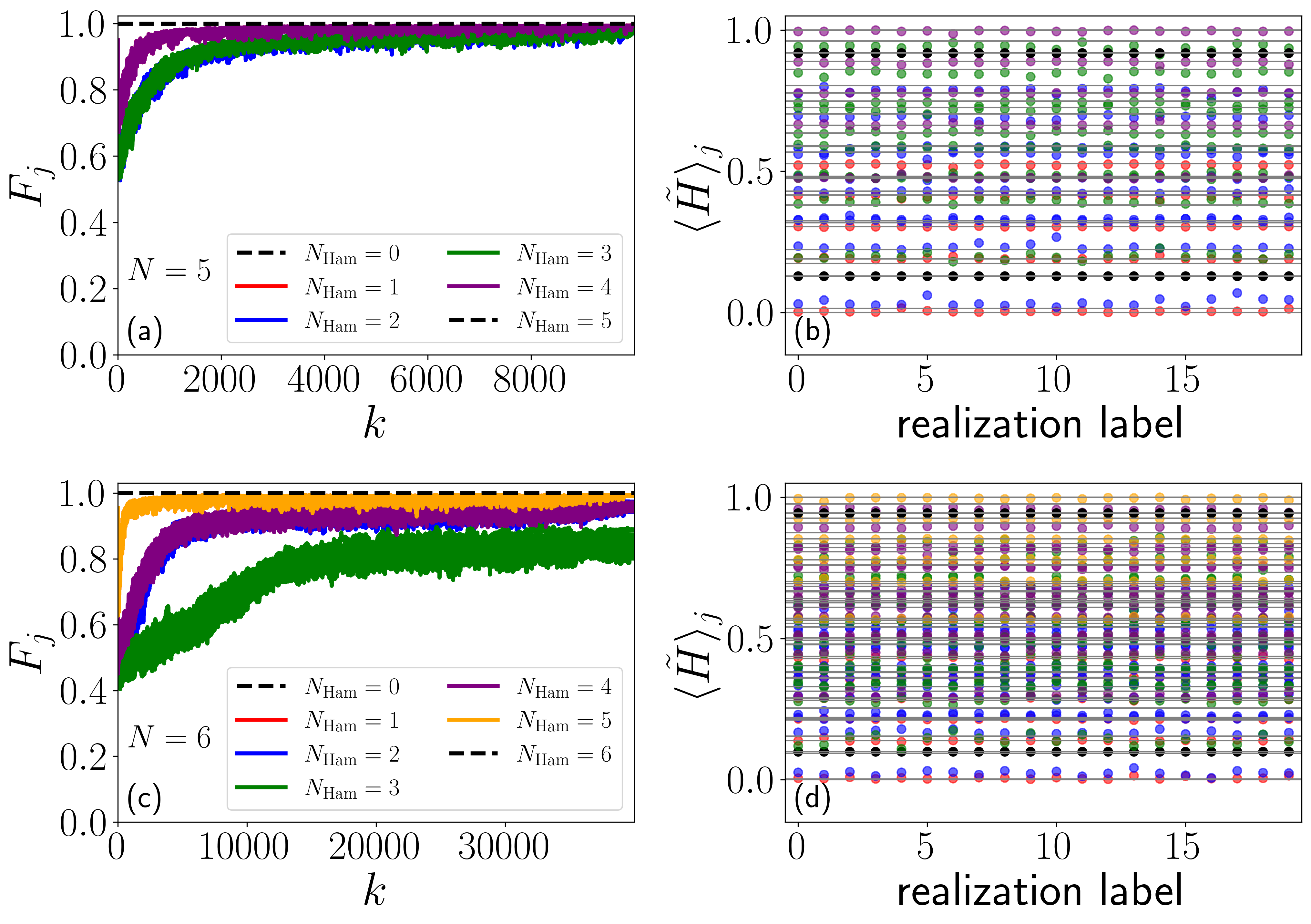}
	\caption{Results of the symmetry-restricted RL algorithm applied to the pairing Hamiltonian. The left panels display, for each computational-basis state, the mean fidelities as a function of the iteration number, while the right panels show, for each state and realization, the dimensionless expectation values of the energies obtained after the stochastic process has converged. Different colors represent the results obtained for the different Hamming-weight sectors. Panels (a) and (b) correspond to the five-qubit system ($N=5$) shown in panels (g) and (h) of Fig.~\ref{fig6}, but with the RL applied independently within each Hamming-weight sector. Panels (c) and (d) show analogous results for a six-qubit system ($N=6$). All parameter values are the same as in Fig.~\ref{fig6}, except for the number of realizations, which is $N_{\mathrm{r}} = 20$ in this case.}
	\label{fig7}
\end{figure}

        Panels (a) and (b) of Fig.~\ref{fig7} show the results obtained from the independent symmetry-restricted calculations corresponding to each Hamming-weight sector: $N_{\mathrm{Ham}} = 0$ ($d = C_5^0=1$), $N_{\mathrm{Ham}} = 1$ ($d = C_5^1=5$), $N_{\mathrm{Ham}} = 2$ ($d = C_5^2=10$), $N_{\mathrm{Ham}} = 3$ ($d = C_5^3=10$), $N_{\mathrm{Ham}} = 4$ ($d = C_5^4=5$), and $N_{\mathrm{Ham}} = 5$ ($d = C_5^5=1$). Since the dimension of each of these sectors is smaller than that of the full Hilbert space, the application of the algorithm becomes less demanding. This figure also confirms our previous interpretation that the different convergence scales observed earlier originate from the different dimensions of the symmetry sectors.

In order to highlight the advantages of this approach, panels (c) and (d) of Fig.~\ref{fig7} show the results of applying the symmetry-restricted RL algorithm to a larger system with $N = 6$ qubits. The fact that some sectors display lower fidelity values does not necessarily indicate a failure of the algorithm. This behavior is analogous to that observed for the four-qubit case in panel (e) of Fig.~\ref{fig6}. In the present case, several eigenenergies are degenerate, and within each degenerate subspace, the eigenstates are not uniquely defined. As a result, the algorithm may converge to linear combinations of these degenerate eigenstates that differ from those obtained by direct diagonalization, leading to reduced fidelities without implying incorrect convergence.

\subsection{Post-selection of ``good'' final states}

An interesting aspect illustrated by the previous figures is the dispersion of the final expectation values of the dimensionless Hamiltonian observed among different realizations. Even though the mean fidelities are high, these expectation values may differ noticeably from the exact eigenenergies from one realization to another. Similarly, the fidelities of the final states also fluctuate between realizations, despite the relatively high mean value.
A natural question that arises is how to determine whether the results obtained in a specific realization are good or bad. The fidelities associated with that realization cannot serve as a criterion, since they could only be computed if the exact eigenstates were already known. Moreover, as previously discussed, low fidelities do not necessarily imply poor convergence to the eigenstates in the presence of degeneracies.
As an alternative, one can consider the energy fluctuations, defined for a given state and realization as
\begin{equation}
\sigma_j=\sqrt{\expval{D_k^\dagger \tilde{H}^2 D_k}{j}-\expval{D_k^\dagger \tilde{H} D_k}{j}^2},
\label{sigma_fluct}
\end{equation}
where $k$ is chosen large enough for the algorithm to have converged. Indeed, this quantity is always non-negative and vanishes only when the state is an exact eigenstate.

\begin{figure}[t]
	\includegraphics[width=0.48\textwidth]{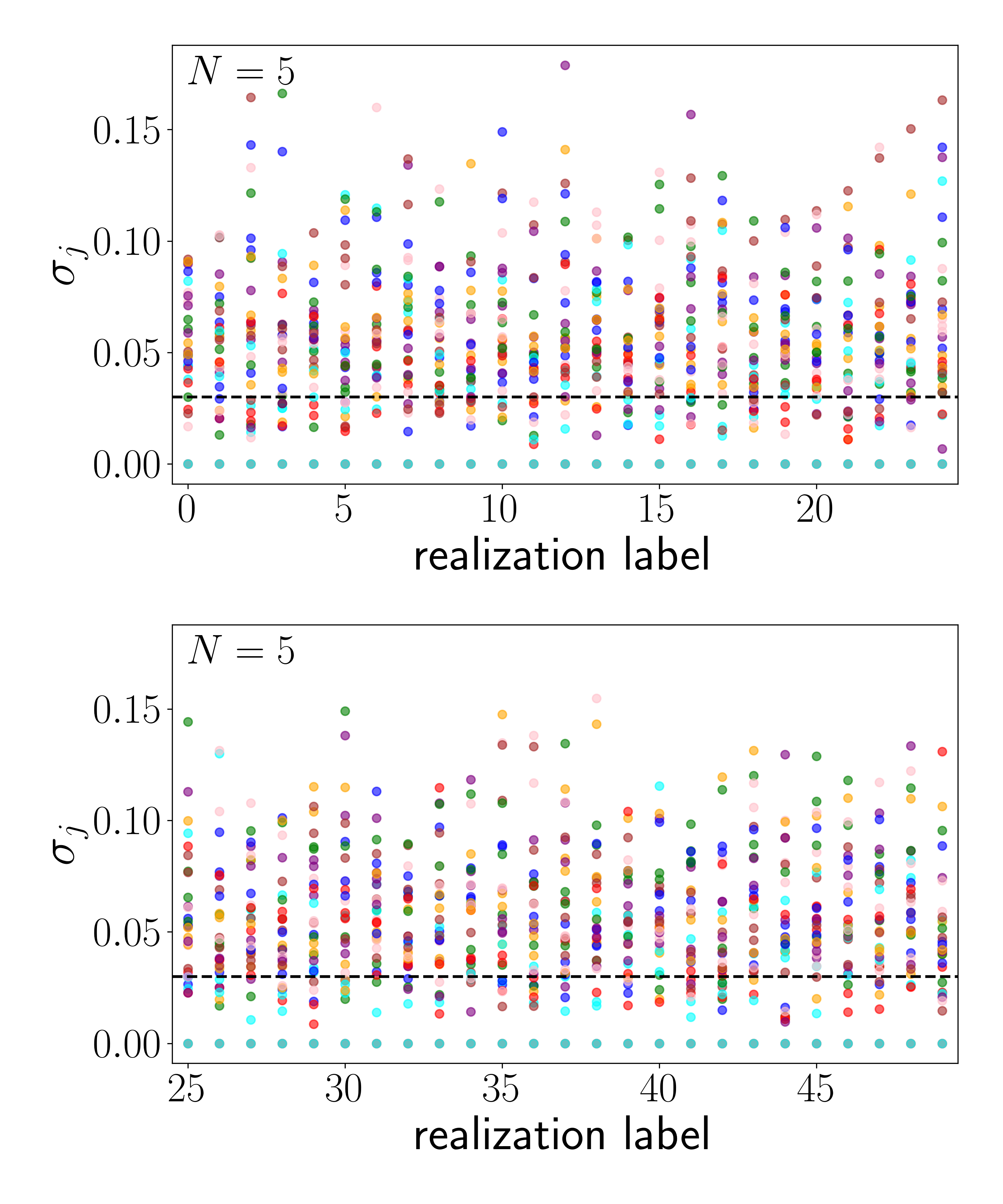}
	\caption{Illustration of the evaluation of the energy fluctuations defined in Eq.~(\ref{sigma_fluct}). This quantity is computed for each final state and plotted as a function of the realization label. The results correspond to the final states obtained for the pairing Hamiltonian with $N=5$ qubits, as in panels (g) and (h) of Fig.~\ref{fig6} and panels (a) and (b) of Fig.~\ref{fig7}. All parameter values are the same as in those figures. The horizontal dashed line indicates the threshold value $\sigma_{\mathrm{th}} = 0.02$, which is used to select the final states shown in Fig.~\ref{fig10}. For clarity, the realizations are divided into two panels: the upper panel shows realizations $0$--$24$, while the lower panel shows realizations $25$--$49$. } 
	\label{fig8}
\end{figure}

\begin{figure}[t]
    \includegraphics[width=0.48\textwidth]{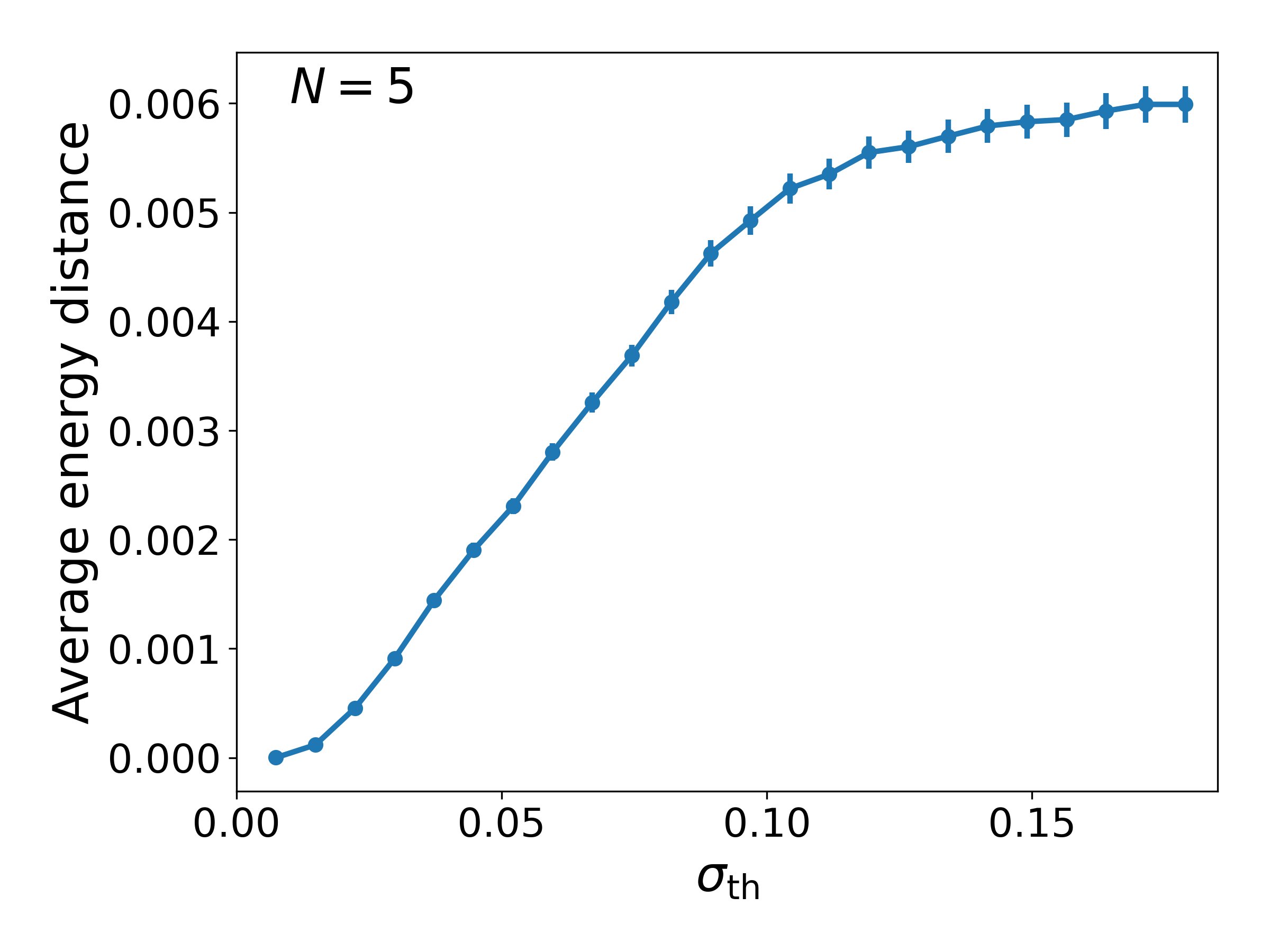}
	\caption{Average distance between the expectation value $\langle \tilde{H} \rangle_j$ and the closest eigenenergy of the dimensionless pairing Hamiltonian, shown as a function of the threshold value $\sigma_{\mathrm{th}}$. The average is computed over all final states and stochastic realizations in Fig.~\ref{fig8} that satisfy the condition $\sigma_j \le \sigma_{\mathrm{th}}$. Error bars indicate statistical deviations across the selected realizations and states.} 
	\label{fig9}
\end{figure}

\begin{figure}[t]
	\includegraphics[width=0.48\textwidth]{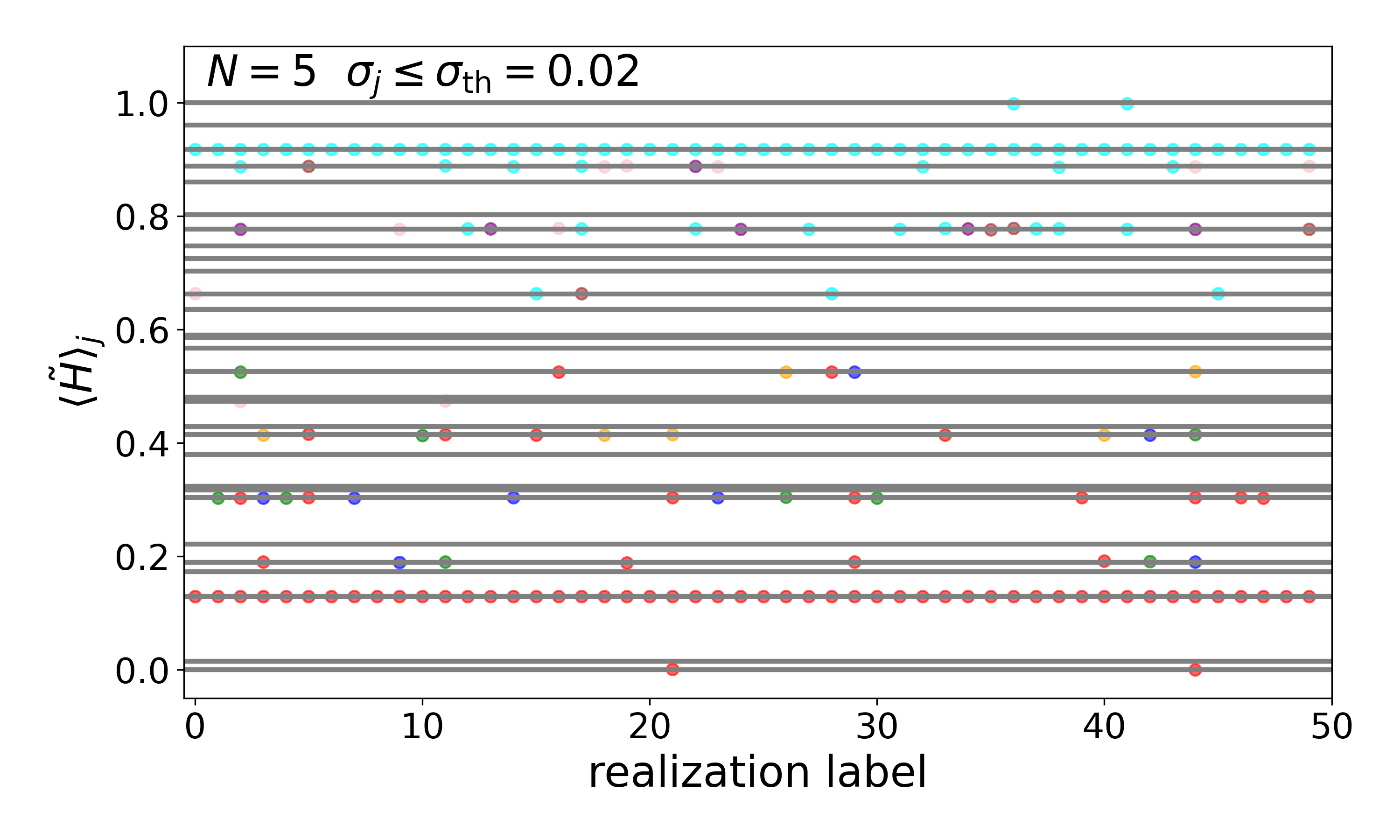}
	\caption{Effect of introducing a fluctuation threshold on the final estimated energies. Only states satisfying $\sigma \le \sigma_{\mathrm{th}}=0.02$ are shown in the energy-versus-realization plot. The results correspond to the $N=5$ pairing Hamiltonian, using the same data as in Fig.~\ref{fig8}. The effect of this post-selection can be appreciated by comparison with panel (f) of Fig.~\ref{fig6}, where all final states are included.} 
	\label{fig10}
\end{figure}

Figure~\ref{fig8} provides a quantitative illustration of the final energy fluctuations computed for the states obtained at the end of the learning process with the pairing Hamiltonian for the case of $N=5$ qubits. The results correspond to the same parameter set used in panels (g) and (h) of Fig.~\ref{fig6} and panels (a) and (b) of Fig.~\ref{fig7}. The figure clearly shows that the fluctuation values exhibit a rather wide dispersion among the different computational-basis states, even within a single realization.

It is to be expected that, for a given state, smaller energy fluctuations indicate a closer convergence to an eigenstate of the Hamiltonian. To further illustrate this, Fig.~\ref{fig9} shows the average distance between the expectation value $\langle \tilde{H} \rangle_j$ and the closest eigenenergy of the dimensionless pairing Hamiltonian as a function of a threshold value $\sigma_{\mathrm{th}}$. The average is taken over all final states and realizations from Fig.~\ref{fig8} that satisfy $\sigma_j \le \sigma_{\mathrm{th}}$, while error bars denote statistical deviations across the selected realizations and states. As observed, the smaller the threshold $\sigma_{\mathrm{th}}$, the smaller the distance between the expectation values and the exact eigenenergies. This observation motivates the introduction of a fluctuation threshold $\sigma_{\mathrm{th}}$ to post-select the most reliable final states.

To illustrate that energy fluctuations can be used as a post-selection criterion to identify ``good'' final states, we present in Fig.~\ref{fig10} the final expectation values of the dimensionless Hamiltonian, retaining only those states whose energy fluctuations fall below a chosen threshold, $\sigma_{\mathrm{th}} = 0.02$. Even when the RL algorithm has not fully converged for all states within a given realization, this selection allows us to isolate the most reliable results. The expectation values obtained from the selected final states show a much better agreement with the exact eigenenergies than those displayed in panel (f) of Fig.~\ref{fig6}, where all final states were included. This confirms that imposing a fluctuation-based selection efficiently filters out poorly converged states while preserving those that faithfully reproduce the true eigenstates of the Hamiltonian.

\section{Conclusions}
\label{sec:conclusions}

In this work, we have introduced a RL-based algorithm designed to identify the unitary transformation that maps the computational basis onto the set of pure states that remain invariant under the action of a given quantum operation. A key feature of this approach is its ability to determine all these states simultaneously and the way it constructs the global unitary transformation iteratively from two-qudit operations guided by a reinforcement strategy. 

To benchmark the algorithm, we considered the problem of finding the eigenstates of specific Hamiltonians. For pseudo-random Hamiltonians involving two- and three-qubit systems, the method achieved high fidelities across multiple stochastic realizations, confirming its robustness and accuracy. We then applied the algorithm to two physical models: the TFIM and the pairing Hamiltonian. In the TFIM case, high fidelities were maintained as the number of qubits increased from two to four, demonstrating the scalability of the approach. For the pairing Hamiltonian, reliable convergence was obtained for systems of up to six qubits, although some states exhibited slightly lower fidelities due to the larger Hilbert-space dimension and the presence of degeneracies.

A remarkable feature of the algorithm is its ability to uncover the underlying symmetries of the system without prior knowledge. This manifests in the appearance of different convergence time scales associated with symmetry sectors of varying sizes. Moreover, by explicitly restricting the learning process to subspaces of fixed Hamming weight, we verified that the algorithm correctly identifies the eigenstates within each symmetry block. Such restriction also makes the computation significantly less demanding, as the effective dimensionality of the learning space is reduced. More generally, these results indicate that the convergence behavior of the algorithm is not determined solely by the total Hilbert-space dimension, but rather by the effective dimension of the space explored during the learning process. In particular, this provides a unified interpretation of the different behaviors observed across the models considered: while in the pairing Hamiltonian the learning dynamics is naturally restricted to symmetry sectors of varying size, in the cases of random Hamiltonians and the TFIM the algorithm explores the full Hilbert space.

We also introduced a post-selection criterion based on the energy fluctuations of the final states, which allows one to identify well-converged (``good'') results even when the RL process has not fully converged for all states. Applying a threshold on these fluctuations efficiently filters out poorly converged states, yielding a set of results that almost perfectly reproduce the exact eigenenergies.

Although in this work we have focused on unitary quantum operations, the formulation of the algorithm only requires that the fixed-point set of the quantum map admit an orthonormal basis of pure states spanning the Hilbert space. When this structural condition is fulfilled, the learning task can still be formulated as the reconstruction of a global unitary transformation. This situation is not restricted to unitary dynamics: certain non-unitary channels, such as phase-damping noise~\cite{NielsenChuang2010}, also satisfy this requirement and are therefore compatible with the present framework.
	In contrast, for quantum channels whose fixed points are mixed states or do not form an orthonormal basis, the present unitary-parametrization strategy would no longer be sufficient, and a different representation of the target manifold would be required. Exploring such generalizations constitutes an interesting direction for future work.
	
The present results should be understood as a proof-of-principle demonstration, as the computational cost grows exponentially with the Hilbert-space dimension and limits the current implementation to moderate system sizes. Within this regime, the interest of the approach lies in the type of learning task it addresses and in its operational setting. In particular, the method enables the global reconstruction of invariant structures of quantum dynamics, providing direct access to symmetry sectors, degeneracy patterns, and spectral properties from dynamical evolution alone, without requiring explicit knowledge of the underlying Hamiltonian. This type of global spectral information is relevant in many-body contexts where the structure of the full eigenbasis plays a central role, such as in the analysis of spectral statistics, thermalization, or excited-state quantum phase transitions. In this sense, the approach is not aimed at large-scale performance, but at providing complementary insight into the structure of quantum dynamics, particularly in scenarios where only dynamical access to the system is available, such as in analog quantum simulators or experimentally implemented many-body systems. From a broader perspective, this measurement-driven and reinforcement-learning-based formulation offers an alternative viewpoint for addressing spectral problems, and may serve as a basis for further methodological developments.

\begin{acknowledgments}
This publication is part of the project PID2022-136228NB-C22, funded by MICIU/AEI/ 10.13039/501100011033 and by ERDF/EU.
Furthermore, this work has been partially supported by the Ministry of Economic Affairs and Digital Transformation of the Spanish Government through the QUANTUM ENIA project call---Quantum Spain project, and by the European Union through the Recovery, Transformation and Resilience Plan---NextGenerationEU within the framework of the ``Digital Spain 2026 Agenda''. It has also been co-financed by the EU, Ministerio de Hacienda y Funci\'on P\'ublica, FEDER, and Junta de Andaluc\'{\i}a (project SOL2024-31833). This work has also received financial support from the CNRS through the AIQI-IN2P3 project. Support from the CNRS-IRN ASTRANUCAP and the Erasmus program project 2024-1-ES01-KA131-HED-000219547 for the visit of M.L.O.-A. and J.C.-P. to D.L. is gratefully acknowledged. This work is part of the HQI initiative (\href{www.hqi.fr}{www.hqi.fr}) and is supported by France 2030 under the French National Research Agency award number ``ANR-22-PNQC-0002''.
\end{acknowledgments}

\appendix

\section{Algorithmic implementation}
\label{appendix}

In this Appendix we present the explicit step-by-step formulation of the learning protocol described in Sec.~\ref{sec:algo}:

\vspace{0.5em}

\begin{algorithmic}

\STATE \textbf{INITIALIZE} $D_1 \leftarrow I$
\STATE \textbf{INITIALIZE} $w_1^{(j,l)} \leftarrow 1$ for all $0 \le j < l \le d-1$

\FOR{$k = 1$ to $k_{\mathrm{M}}$}

\STATE \textbf{PREPARE} $d$ registers in the computational basis
\STATE \hspace{1em} $\rho^{(j)} \leftarrow \ketbra{j}{j}$ for $j=0,\dots,d-1$

\STATE \textbf{APPLY} global unitary transformation
\STATE \hspace{1em} $\rho^{(j)}  \leftarrow D_k \rho^{(j)}  D_k^\dagger$ for all $j$

\STATE \textbf{APPLY} quantum operation
\STATE \hspace{1em} $\rho^{(j)} \leftarrow \mathcal{E}(\rho^{(j)} )$ for all $j$

\STATE \textbf{UNDO} global unitary
\STATE \hspace{1em} $\rho^{(j)} \leftarrow D_k^\dagger \rho^{(j)}  D_k$ for all $j$

\STATE \textbf{MEASURE} in the computational basis
\STATE \textbf{RECORD} outcomes $m_k^{(j)}$ for all $j$

\FOR{each pair $(j,l)$ with $0 \le j < l \le d-1$}

\IF{$m_k^{(j)} \neq l$ and $m_k^{(l)} \neq j$}

\STATE \textbf{UPDATE} $w_{k+1}^{(j,l)} \leftarrow r^2 w_k^{(j,l)}$
\STATE \textbf{SET} $D_k^{(j,l)} \leftarrow I$

\ELSIF{exactly one of $m_k^{(j)} = l$, $m_k^{(l)} = j$}

\STATE \textbf{UPDATE} $w_{k+1}^{(j,l)} \leftarrow \min(rp\, w_k^{(j,l)}, 1)$
\STATE \textbf{SAMPLE} $\alpha_k^{(j,l)}, \beta_k^{(j,l)}, \gamma_k^{(j,l)}$
\STATE \hspace{1em} uniformly in $[-\pi w_k^{(j,l)}, \pi w_k^{(j,l)}]$
\STATE \textbf{CONSTRUCT} $D_k^{(j,l)}$ according to Eq.~\eqref{eq:random-rot}

\ELSE

\STATE \textbf{UPDATE} $w_{k+1}^{(j,l)} \leftarrow \min(p^2 w_k^{(j,l)}, 1)$
\STATE \textbf{SAMPLE} $\alpha_k^{(j,l)}, \beta_k^{(j,l)}, \gamma_k^{(j,l)}$
\STATE \hspace{1em} uniformly in $[-\pi w_k^{(j,l)}, \pi w_k^{(j,l)}]$
\STATE \textbf{CONSTRUCT} $D_k^{(j,l)}$ according to Eq.~\eqref{eq:random-rot}

\ENDIF

\ENDFOR

\STATE \textbf{UPDATE} global unitary
\STATE \hspace{1em} $D_{k+1} \leftarrow D_k \prod_{j<l} D_k^{(j,l)}$

\IF{$\max_{j<l} \big(w_{k+1}^{(j,l)}\big)< w_{\mathrm{th}}$}
\STATE \textbf{BREAK}
\ENDIF

\ENDFOR

\STATE \textbf{OUTPUT} learned unitary transformation $D_k$

\end{algorithmic}

\section*{Data Availability}
The data supporting the findings of this study, including a documented code example illustrating the implementation used in this work, are available from the corresponding author upon request.

\bibliography{literature_Denis,biblio-QC}

\end{document}